\documentclass[a4paper,12pt, twocolumn]{article}
\usepackage{mathptmx} 
\usepackage{graphicx}
\usepackage{amsmath}
\usepackage{geometry}
\usepackage{fancyhdr}
\usepackage{multicol}
\usepackage{titlesec}
\usepackage[english]{babel} 
\usepackage{hyphenat}
\usepackage[numbers,sort&compress]{natbib} 
\usepackage{wasysym}
\usepackage{nicefrac}
\usepackage{textcomp}
\usepackage{gensymb}
\usepackage[T1]{fontenc} 
\usepackage{pifont} 
\usepackage{stfloats} 
\usepackage{multirow}
\usepackage{caption}
\usepackage{comment}
\usepackage{subcaption}
\usepackage{color}
\usepackage{moresize}
\usepackage[colorlinks=true, linkcolor=black, citecolor=black, urlcolor=blue]{hyperref}
\usepackage{microtype}
\usepackage{physics}
\usepackage{array}
\usepackage{tabularx} 
\usepackage{longtable}

\DeclareUnicodeCharacter{202F}{\,}

\renewcommand{\baselinestretch}{1.0}

\makeatletter
\renewcommand{\NAT@open}{[} 
\renewcommand{\NAT@close}{]} 
\makeatother

\makeatletter
\renewenvironment{abstract}{
    \noindent
    \textbf{} %
    \quotation
}{\endquotation}
\makeatother

\titleformat{\section}{\normalfont\normalsize\bfseries}{}{0pt}{}
\titlespacing*{\section}{0pt}{12pt}{2pt}

\titleformat{\subsection}{\normalfont\normalsize\bfseries}{}{0pt}{}
\titlespacing*{\subsection}{0pt}{2pt}{2pt}

\titleformat{\subsubsection}{\normalfont\normalsize\bfseries\itshape}{}{0pt}{}
\titlespacing*{\subsubsection}{0pt}{4pt}{2pt}

\title{Picogram-Level Nanoplastic Analysis with Nanoelectromechanical System Fourier Transform Infrared Spectroscopy: NEMS-FTIR}

\author{
    \fontsize{11pt}{11pt}\selectfont Jelena~Timarac--Popovi\'c\textsuperscript{1,2*}, Johannes~Hiesberger\textsuperscript{2}, Eldira~\v{S}esto\textsuperscript{1}, Niklas~Luhmann\textsuperscript{2},\\ 
    \fontsize{11pt}{11pt}\selectfont Ariane~Giesriegl\textsuperscript{1}, Hajrudin~Be\v{s}i\'c\textsuperscript{1,2}, Josiane~P.~Lafleur\textsuperscript{2}, and Silvan~Schmid\textsuperscript{1}\footnote{\mbox{Correspondence e-mail address:} \mbox{jelena.popovic@tuwien.ac.at}, silvan.schmid@tuwien.ac.at} \\[2mm]
    \fontsize{10pt}{10pt}\selectfont \textsuperscript{1}\textit{TU Wien, Institute of Sensor and Actuator Systems, Gusshausstrasse 27--29, 1040 Vienna, Austria.} \\[-0.5mm]
    \fontsize{10pt}{10pt}\selectfont \textsuperscript{2}\textit{Invisible--Light Labs GmbH, Taubstummengasse 11, 1040 Vienna, Austria.} \\[1mm]
    \fontsize{10pt}{10pt}\selectfont \textit{*Corresponding authors: jelena.popovic@tuwien.ac.at, silvan.schmid@tuwien.ac.at}\\[2mm]
   }

\date{\vspace{-11pt} \fontsize{10pt}{12pt}\selectfont (Dated: \today) \vspace{0pt}}

\begin{document}

\twocolumn[
    \begin{@twocolumnfalse}
    \maketitle
    \vspace{-30pt}
    \begin{abstract}
    \fontsize{10pt}{12pt}\selectfont 
    \noindent We present a photothermal infrared spectroscopy-based approach for the chemical characterization and quantification of nanoplastics. By combining the high sensitivity of nanoelectromechanical systems (NEMS) with the wide spectral range and ubiquity of commercially available Fourier transform infrared (FTIR) spectrometers, NEMS-FTIR offers a time-efficient and cryogen-free option for the rapid, routine analysis of nanoplastics in aqueous samples. Polypropylene, polystyrene, and polyvinyl chloride nanoplastics with nominal diameters ranging from 54 to 262~nm were analyzed by NEMS-FTIR with limits of detection ranging from 101~pg to 353~pg, one order of magnitude lower than values reported for pyrolysis–gas chromatography–mass spectrometry of nanoplastics. The absorptance measured by NEMS-FTIR could be further converted to absolute sample mass using the attenuation coefficient, as demonstrated for polystyrene. Thanks to the wide spectral range of NEMS–FTIR, nanoplastic particles from different polymers could be readily identified, even when present in a mixture. The potential of NEMS-FTIR for the analysis of real samples was demonstrated by identifying the presence of nanoplastics released in water during tea brewing. Polyamide leachates in the form of fragments and smaller oligomers could be identified in the brewing water without sample pre-concentration, even in the presence of an organic matrix. Accelerated aging of the nylon teabags under elevated temperature and UV radiation showed further release of polyamide over time.
    \end{abstract}
    \noindent \fontsize{10pt}{12pt}\selectfont\textbf{Keywords:} nanoplastics, nanoparticles, NEMS-FTIR, Fourier transform infrared spectroscopy, photothermal sensing, nanoelectromechanical system, nylon teabag.
    \vspace{25pt}
    \end{@twocolumnfalse}
]

\section{INTRODUCTION}
Nanoplastics have become ubiquitous and pose significant environmental and health risks due to their high reactivity, potential to transport pollutants, and widespread dispersion across ecosystems. \cite{dick2024adsorptionofantibiotics, amaral2020ecologyplastisphere, koelmans2014leaching, wright2013impactmarineorganisms, wang2021NPsfate, fred2020interaction, horton2018transport} The ability of nanoplastics to penetrate deep into tissues underscores the need for their routine chemical characterization and monitoring. \cite{ Ali2024impactonhumanhealth}
Sampling of nanoplastic particles is particularly challenging due to their nanoscale dimensions and low concentrations, which often exceed the capabilities of many analytical techniques.  \cite{Käppler2016MPsanalysisFTIRRaman, Adhikari2022mini-review, Caldwell2022detectionmethods} While state-of-the-art techniques allow for routine microplastic detection and identification, nanoplastics, on the other hand, remain a challenge to this day.

Mass spectrometry-based techniques can be used for the bulk chemical identification of nanoplastics. \cite{Adhikari2022mini-review, Caldwell2022detectionmethods, Cai2021NPsprogress} Pyrolysis–gas chromatography–mass spectrometry (Py-GC/MS) has been extensively applied to detect various types of nanoplastics, including polystyrene (PS), polypropylene (PP), polycarbonate (PC), polyethylene terephthalate (PET), polymethyl methacrylate (PMMA), polyvinyl chloride (PVC), polyethylene (PE), and Nylon 6/66, with reported limits of detection (LoDs) as low as 1~ng/L. \cite{Xu2022PyGC-MS, Xu2023PyGC-MS, Li2025PyGC-MS} Several studies report Py-GC/MS LoDs in the low-nanogram range for polymer analysis in aqueous samples. \cite{Hasager2024PyGC-MS, li2024PyGC-MS, Fischer2019PyGC-MS}
A combination of atomic force microscope-IR (AFM-IR) spectroscopy and Py-GC/MS was also used to detect and quantify PE and PVC particles (20–1000~nm), with LoDs of 280~ng for PE and 1.54~$\mu$g for PVC. \cite{Li2024AFM-IRandPyGC-MS} Alternative approaches, such as thermal desorption-proton transfer reaction-mass spectrometry (TD-PTR-MS), demonstrated remarkable sensitivity, with Materi\'c \textit{et al}.~\cite{Materic2020TD-PTR-MS} detecting pure 1~$\mu$m PS particles with an LoD of 340~pg, allowing detection of these particles in environmental samples as small as 1~mL without preconcentration. While mass spectrometry-based techniques provide excellent capabilities, their complexity and cost hinder their use in routine environmental monitoring. \cite{Caldwell2022detectionmethods, Cai2021NPsprogress, Zhang2024MS, Li2025PyGC-MS, Jing2025PyGC-MS}

FTIR spectroscopy with its broad spectral range allows functional group chemical identification of unknown samples and is one of the most extensively used methods in microplastic analysis. \cite{Xie2024IRreview,Primpke2020Review}  
Commonly used attenuated total reflectance - FTIR (ATR-FTIR) spectroscopy is a contact bulk analysis technique with a detection limit in the order of 10~ng. \cite{Casci2019thinpolymericfilms} 
FTIR microscopy methods, such as µ-FTIR scanning microscopy and focal plane array-FTIR (FPA-FTIR) microscopy, offer a contact-less FTIR solution, reaching LoDs as low as 450~pg when used with a cryogenically cooled MCT array. \cite{lux2022limit}  However, diffraction limits the reliable analysis of single particles with sizes below 10~$\mu$m. \cite{Xie2024IRreview, Primpke2020Review}

Quantum cascade laser infrared (QCL-IR) spectroscopy provides higher sensitivity, albeit with a narrower spectral range and additional issues with coherence artifacts \cite{Xie2024IRreview}. Although QCLs cover the fingerprint region relevant for polymer identification and provide stronger spectral signals due to their high source intensity, they typically have higher noise levels and limited spectral coverage, missing informative regions such as the C–H stretching vibrations around 3000~cm$^{-1}$ and the low-wavenumber region around 700~cm$^{-1}$. Full-range spectra, as enabled by FTIR, facilitates chemometric analysis and spectral deconvolution, which can help identify and quantify nanoplastics in multicomponent samples. While such methods can also work within narrower spectral windows, having access to broader spectral information can make differentiation easier and more reliable. QCL-IR microscopy is a complementary method to FTIR microscopy and is therefore sometimes offered in tandem.\footnote{HYPERION II from Bruker Corporation} Primpke \textit{et al}.~\cite{Primpke2020QCL-IR} demonstrated that QCL-based microscopy can detect polymer particles as small as 1.4~$\mu$m. All standard IR spectroscopy techniques suffer from Mie scattering-related spectral artifacts when measuring particulate samples. \cite{bassan2009resonant}

Scattering, diffraction, and coherence issues can be reduced by QCL-based photothermal scanning methods, such as optical–photothermal IR (O-PTIR) spectroscopy or AFM-IR microscopy. O-PTIR has proven effective for the imaging with sub-micrometer lateral resolution of individual PS nanoparticles as small as 250~nm in mammalian tissues \cite{duswald2024O-PTIR} and identification of plastics (>~600~nm) containing polysiloxanes and imides, originating from the steam disinfection of silicone baby teats. \cite{su2022O-PTIR} However, O-PTIR is sensitive to the relative focus alignment between the pump and probe beams, and to the axial signal distribution, which can introduce artifacts in depth-resolved measurements.\cite{Holub2025O-PTIRartifacts} Even better lateral resolution and single nanoparticle analysis can be obtained by AFM-IR. \cite{dazzi2017afm, Xie2024IRreview} However, AFM-IR faces challenges with low signal-to-noise ratio (SNR) and slow imaging speed. \cite{Xie2024IRreview}

Raman spectroscopy has the advantage of probing at visible wavelengths with low interference from water and, similar to O-PTIR, allows for the analysis of plastic particles down to sub-micrometer sizes, with spatial resolution sufficient to resolve and chemically identify individual particles. \cite{sobhani2020identification} Surface-enhanced Raman spectroscopy (SERS) has emerged as a promising technique. Zhou \textit{et al}.~\cite{Zhou2021SERS} and Hu \textit{et al}.~\cite{Hu2022SERS} reported LoDs as low as 5~$\mu$g/mL for PS particles as small as 50~nm using conventional SERS substrates.
Hyperspectral stimulated Raman scattering (SRS) microscopy has been demonstrated as a powerful tool for the chemical imaging of nanoplastics, enabling identification and classification of particles down to 130~nm. \cite{qian2024SRS}
However, SERS suffers from issues such as spectral artifacts and reliance on carefully engineered substrates, limiting its versatility. \cite{Caldwell2022detectionmethods, pei2023review} SRS microscopy, despite offering rapid detection and high sensitivity, relies on complex and expensive instrumentation. \cite{asamoah2021review} Table~\ref{tab:Comparison between analytical techniques} summarizes the advantages and disadvantages of the most commonly used analytical techniques for nanoplastic identification.

Although existing analytical techniques such as Py-GC/MS, QCL-IR, O-PTIR, or SERS offer high sensitivity and specificity, they remain impractical for routine nanoplastics monitoring, particularly for water suppliers and official control laboratories.~\cite{JRC2024Review}

Nanoelectromechanical systems IR spectroscopy (NEMS-IR) is a technique based on the photothermal effect that has been introduced in recent years. It employs NEMS resonators, or chips, as its key component. \cite{Nik2023thermaldesorption, Kurek2017pharmaceuticalcompounds} The NEMS chip can be used simultaneously as a sample carrier and detector. The NEMS chips, typically consisting of a pre-stressed few-nanometer-thin silicon nitride (SiN) membrane, are robust and allow for a wide variety of sampling techniques, \cite{west2023photothermal} including direct aerosol collection \cite{schmid2013real, Kurek2017pharmaceuticalcompounds, Nik2023thermaldesorption} and drop casting.  When a sample deposited on the NEMS chip's surface absorbs IR light, local heating occurs, causing thermal expansion and tensile stress reduction. The corresponding frequency detuning of the NEMS chip is proportional to the absorbed power, and the frequency shift can be monitored using a closed-loop oscillation scheme in combination with a frequency counter, as demonstrated by Be\v{s}i\'c \textit{et al}.~\cite{besic2023schemes, besic2023FC-SSO} Unlike in imaging techniques, the entire sensing area of the NEMS chip is illuminated by the probing IR light. The resulting IR spectrum reveals the bulk properties of the sample.  NEMS-IR has shown picogram sensitivity at room temperature over a large spectral range, from the ultraviolet to the far-IR. \cite{west2023photothermal}  This technique has been used to analyze polymer nanoparticles, \cite{Yamada2013polymericNPS, besic2024NEMS-IR} pharmaceutical compounds, \cite{Kurek2017pharmaceuticalcompounds, Samaeifar2019pharmaceuticalcompounds} polymer micelles, \cite{Andersen2016micelles} thin polymer films \cite{Casci2019thinpolymericfilms} and explosives. \cite{Biswas2014explosives} NEMS-IR with \textit{in-situ} thermal desorption (TD) analysis has also been demonstrated for separating simple analyte mixtures by Luhmann \textit{et al}.~\cite{Nik2023thermaldesorption} 

\begin{figure*}[h!]
\centering
\includegraphics[width=\textwidth]{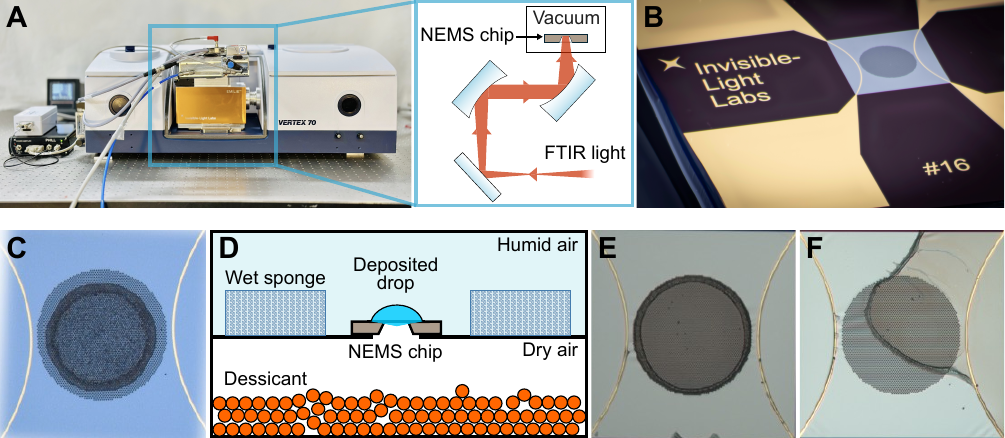}
\caption{\textbf{Experimental setup for NEMS-FTIR analysis.} (\textbf{A})~An FTIR spectrometer (Vertex 70, Bruker Corporation, MA, USA) equipped with the nanoelectromechanical IR analyzer (EMILIE$^{\text{TM}}$, Invisible-Light Labs GmbH, Austria) for NEMS-FTIR analysis. The inset illustrates the optical path of the IR beam from the spectrometer output to the NEMS chip. (\textbf{B})~A NEMS sampling and sensing chip (Invisible-Light Labs GmbH, Austria). The central square membrane on the NEMS resonator is made of SiN and has a lateral size of $1\times 1$~mm$^2$ and a thickness of $\sim$ 50~nm. It features a central circular perforated area with a diameter of approximately 600~$\mu$m, consisting of 6~$\mu$m holes spaced 3~$\mu$m apart. (\textbf{C})~20~nL drop of PS dispersion, corresponding to a deposited mass of 15~ng, deposited and confined within the perforated membrane area using a piezoelectric nanodroplet dispenser. (\textbf{D})~Schematic of drop casting combined with the pervaporation method, where a humidity gradient causes solvent evaporation through the perforation to collect the sample in the membrane's central region. (\textbf{E}) 500~nL drop of PS dispersion, corresponding to a deposited mass of 15~ng, drop casted with a micropipette on the membrane area and dried using the pervaporation method, and (\textbf{F}) without using the pervaporation method.}
\label{fig: Figure 1}
\end{figure*}

NEMS-IR has only been used for dispersive spectroscopy, typically relying on narrow-band tunable QCLs. Optomechanical metalized cantilevers have previously been used as photothermal detectors in combination with FTIR sources, \cite{tetard2011optomechanical} but their lower sensitivity, \cite{schmid2023fundamentals} reliance on optical readout, and limited surface area for sample collection have constrained their practical application.  This paper introduces NEMS-FTIR spectroscopy, interfacing NEMS with FTIR spectrometers as a light source (see Fig.~\ref{fig: Figure 1}A).

By combining the broad spectral coverage, wide availability, and ease of use of commercially available FTIR spectrometers with the high sensitivity of nanomechanical photothermal detection using NEMS resonators, NEMS–FTIR enables measurements with picogram-level detection limits -- without the need for cryogenic cooling. Because the readout is purely photothermal, NEMS–FTIR is inherently immune to common IR spectral artifacts, including Mie-scattering distortions \cite{Dazzi2013scattering}, ATR-related spectral anomalies \cite{bradley2018ATRpathlength,  mayerhofer2021ATR, grdadolnik2002ATR, miseo2025ATR, boulet2010ATR}, and coherence issues.\cite{Chalmers2001IRartifacts}

The responsivity of the NEMS chip’s membrane is highest at its center, where a circular microperforation is located (see Fig.~\ref{fig: Figure 1}B). Consequently, the signal is maximized when the deposited analyte is confined to this region. Previous studies have achieved such localization using an aerosol-based deposition method.\cite{schmid2013real, Kurek2017pharmaceuticalcompounds} In this approach, liquid samples are nebulized and the resulting aerosol particles, containing the analyte, are drawn through the central perforation of the NEMS chip, where they are collected with high efficiency. \cite{Nik2023thermaldesorption,Kurek2017pharmaceuticalcompounds} Although this method yields a uniform coverage of the entire perforated area (see Fig.~S1B), the particle-size-dependent collection efficiency complicates quantitative analysis. To address this limitation, we introduce two drop-casting deposition methods that ensure 100\% retention of non-volatile analytes, thereby enabling quantitative measurements.

Combined with these quantitative deposition strategies, the straightforward sampling of nanoparticle dispersions directly onto the disposable NEMS chip makes NEMS–FTIR a powerful bulk analysis method for the routine detection and characterization of nanoplastics. 

\clearpage
\onecolumn
\setlength{\LTcapwidth}{\linewidth}
{\footnotesize
\begin{longtable}{
  >{\raggedright\arraybackslash}p{2.5cm}
  >{\raggedright\arraybackslash}p{5.2cm}
  >{\raggedright\arraybackslash}p{5.2cm}
  >{\raggedright\arraybackslash}p{1.8cm}
}

\caption{\textbf{Comparison of different analytical techniques for nanoplastic identification.} Advantages and disadvantages of the most commonly used techniques for identifying nanoplastics.
}
\label{tab:Comparison between analytical techniques}
\\
\hline
\textbf{{\raggedright Technique}} & 
\textbf{{\raggedright Advantages}} & 
\textbf{{\raggedright Disadvantages}} & 
\textbf{{\raggedright References}} \\[0.3em]
\hline
\endfirsthead

\hline
\textbf{{\raggedright Technique}} & 
\textbf{{\raggedright Advantages}} & 
\textbf{{\raggedright Disadvantages}} & 
\textbf{{\raggedright References}} \\[0.3em]
\hline
\endhead

\multicolumn{4}{l}{\textit{\textbf{Chemical imaging techniques}}} \\
\hline
\textbf{µ-FTIR / FPA-FTIR microscopy}
& Broad spectral range \newline
450~pg LoD with cooled MCT array
& Diffraction-limited; unreliable below 10~$\mu$m particles \newline
High instrumentation cost
& \cite{Xie2024IRreview, Primpke2020Review, lux2022limit}\\
\hline
\textbf{QCL-IR microscopy}
& High signal-to-noise ratio \newline
Detects particles down to 1.4~$\mu$m \newline
& Narrow spectral range \newline
Higher noise than FTIR \newline
Coherence artifacts \newline
High instrumentation cost 
& \cite{Xie2024IRreview, Primpke2020QCL-IR}\\
\hline
\textbf{O-PTIR}
& Avoids Mie/diffraction artifacts \newline
High signal-to-noise ratio  \newline
Sub-micron IR imaging down to 250~nm
& Narrow spectral range \newline
Long acquisition times  \newline
High instrumentation cost
& \cite{duswald2024O-PTIR, su2022O-PTIR, Xie2024IRreview}\\
\hline
\textbf{AFM-IR}
& High spatial resolution ($\sim$20~nm) \newline
Effective for particles from $\sim$20~nm up to 1~$\mu$m 
& Narrow spectral range \newline
Long acquisition times \newline
Low signal-to-noise ratio
& \cite{dazzi2017afm, Xie2024IRreview, Adhikari2022mini-review, Caldwell2022detectionmethods, Primpke2020Review}\\
\hline
\textbf{Raman microscopy}
& High spatial resolution\newline
Low water interference \newline
Suitable for particles 1-100~$\mu$m 
& Fluorescence interference \newline
Long acquisition times \newline
Low sensitivity for some polymers 
& \cite{Caldwell2022detectionmethods, Adhikari2022mini-review, Cai2021NPsprogress, Käppler2016MPsanalysisFTIRRaman}\\
\textbf{SERS microscopy}
& Very high sensitivity \newline
Detects particles down to 50~nm \newline
5~$\mu$g/mL LoD
& Requires engineered substrates \newline
Long acquisition times \newline
Spectral artifacts \newline
Polymer-dependent enhancement
&\cite{Zhou2021SERS, Hu2022SERS, yang2022SERS, Caldwell2022detectionmethods, pei2023review}\\
\hline
\textbf{SRS microscopy}
& Rapid imaging with high contrast \newline
Detects particles down to $\sim$130~nm
& High instrumentation cost
&\cite{qian2024SRS, asamoah2021review, Primpke2020Review}\\
\hline
\multicolumn{4}{l}{\textit{\textbf{Bulk techniques}}} \\
\hline
\textbf{Py-GC/MS}
& Ability to detect polymers and plastic aditives \newline
Applicable to complex matrices \newline
1-10~ng LoDs
& Destructive technique \newline
Reduced chemical specificity for micro- and nanoplastics \newline
No standardized protocols \newline
Background interference \newline
Small injection volume \newline
High instrumentation cost
& \cite{Hasager2024PyGC-MS, Fischer2019PyGC-MS, Caldwell2022detectionmethods, Cai2021NPsprogress, Adhikari2022mini-review, Okoffo2024PyGC-MS, Xu2022PyGC-MS, Xu2023PyGC-MS, li2024PyGC-MS, Primpke2020Review, Li2025PyGC-MS, Jing2025PyGC-MS}\\
\hline
\textbf{TD-PTR-MS}
& High sensitivity for small sample sizes \newline
340~pg LoD
& Destructive technique \newline
Complex data analysis \newline
High instrumentation cost
& \cite{Caldwell2022detectionmethods, Materic2020TD-PTR-MS, Cai2021NPsprogress}\\
\hline
\textbf{ATR-FTIR}
& Widely available and easy to use \newline
$\sim$10~ng LoD
& Physical contact with the sample \newline
Risk of cross-contamination \newline
Prone to spectral artifacts
& \cite{Xie2024IRreview, Primpke2020Review, Casci2019thinpolymericfilms, Käppler2016MPsanalysisFTIRRaman}\\
\hline
\textbf{NEMS-FTIR}
& No lower particle-size limit \newline
Minimal spectral artifacts \newline
Compatible with transmission FTIR spectral libraries \newline
Allows downstream analysis by complementary methods such as SEM and O-PTIR \newline
Intrinsic SiN internal standard \newline
101-353~pg LoD
& Requires disposable sensing chips \newline
Upper particle-size limit (circa 5~$\mu$m)\newline
Measurements under vacuum preclude the analysis of volatile compounds 
&\textbf{This work}\\
\hline

\end{longtable}
}

\twocolumn

Many studies have analyzed the release of micro- and nanoplastic particles from teabags during tea preparation using different analytical methods such as scanning transmission X-ray microscopy, \cite{foetisch2022teabags} IR photothermal heterodyne microscopy, \cite{kniazev2021teabags} Raman imaging, \cite{mei2022teabags, Yue2024teabag, Yousefi2024teabag, Busse2020teabagoligomers}, liquid chromatography-MS-based techniques, \cite{Tsochatzis2020teabagoligomers, Busse2020teabagoligomers, Kappenstein2018teabagoligomers}, and ATR-FTIR spectroscopy. \cite{banaei2024teabag, hernandez2019teabags} To avoid the complexity of the tea leaves matrix, these investigations were commonly done with emptied teabags, which simplifies sample preparation \cite{Fard2025teabag, kniazev2021teabags, mei2022teabags, banaei2024teabag, hernandez2019teabags, Tsochatzis2020teabagoligomers}. Several experimental strategies deviate notably from everyday brewing conditions, for example by using much higher teabag-to-water ratios \cite{hernandez2019teabags, banaei2024teabag} or involving sample up-concentration steps such as ultracentrifugation, \cite{banaei2024teabag} density separation {and concentration by ultrafiltration, \cite{foetisch2022teabags} or evaporation drying to obtain a powder. \cite{hernandez2019teabags}. 

In contrast, we demonstrate here that NEMS–FTIR can chemically identify nylon-based PA leachates in the brewing water from a single teabag. NEMS-FTIR analysis of the leachate required just a few nL sample aliquot and did not require any pre-concentration steps, even in the presence of the organic tea leaves matrix.  The absence of size limitations in NEMS-FTIR allowed for the identification of both nylon polymer fragments as well as smaller nylon oligomers leached from the nylon teabags. Morphological information could be obtained by imaging samples deposited on NEMS chips directly by scanning electron microscopy (SEM). The LoD of NEMS-FTIR was determined for three model nanoplastics -- PS, PP, and PVC -- and the measured absorbance was quantitatively related to the actual mass of PS particles deposited on the NEMS chips.

\section{RESULTS AND DISCUSSION}
\subsection{Drop casting of aqueous nanoplastics dispersions on the NEMS chips}
To enable quantitative measurements, two drop casting methods were used in this work to deposit and confine nanoplastic samples from a water dispersion at the perforated center of the NEMS chip. 1)~Nanoliter droplets were deposited with a piezoelectric nanodroplet dispenser. Following a 2-minute drying time for a 20~nL droplet, the particles were fully confined within the perforated area of the NEMS chips, as shown in Fig.~\ref{fig: Figure 1}C.
The coffee ring diameter (see Fig.~S2), corresponding to the dried droplet footprint on the membrane, was determined from six independent measurements, yielding an average size of 252$\pm$23~$\mu$m for 20~nL droplets, confirming the reproducibility of droplet size.
2)~For very low concentration samples, droplets as large as 500~nL were deposited on the NEMS chip. These droplets dispensed \textit{via} a micropipette have a diameter larger than the 600~$\mu$m diameter of the perforated area of the membrane. Drying using a combination of permeation and evaporation, or pervaporation, as schematically shown in Fig.~\ref{fig: Figure 1}D, was used to collect the analytes in the sensing area. During the pervaporation process, a controlled humidity gradient was created across the membrane to drive solvent evaporation preferentially through the membrane perforation. As a result, solutes and particulates deposited from large droplets consistently dried in the center of the membrane, as shown in Fig.~\ref{fig: Figure 1}E. By contrast, without pervaporation, the analyte dried across the entire chip (see Fig.~\ref{fig: Figure 1}F). After drying, further droplets can be added to concentrate the sample. A 500~nL drop dried within approximately 30~minutes using the pervaporation method.

\subsection{Single nanoplastic dispersions}
Aqueous dispersions of three different polymers, PS (nominal diameter: 100~nm), PP (nominal diameter: 54~nm), and PVC (nominal diameter: 262~nm), were deposited on individual NEMS chips using piezoelectric nanodroplet dispenser. 
Fig.~\ref{fig: Figure 2_PS}A shows PS nanoparticles after deposition onto the central circular perforation of the NEMS chips (see Fig.~\ref{fig: Figure 1}B). Corresponding scanning electron microscopy (SEM) images of PP and PVC nanoparticles deposited on NEMS chips are provided in the Supporting Information in Fig.~S3. These SEM images highlight the generally spherical morphology and relatively uniform size distributions of the model nanoplastic particles studied. 

\begin{figure*}[h!]
\centering
\includegraphics[width=\textwidth]{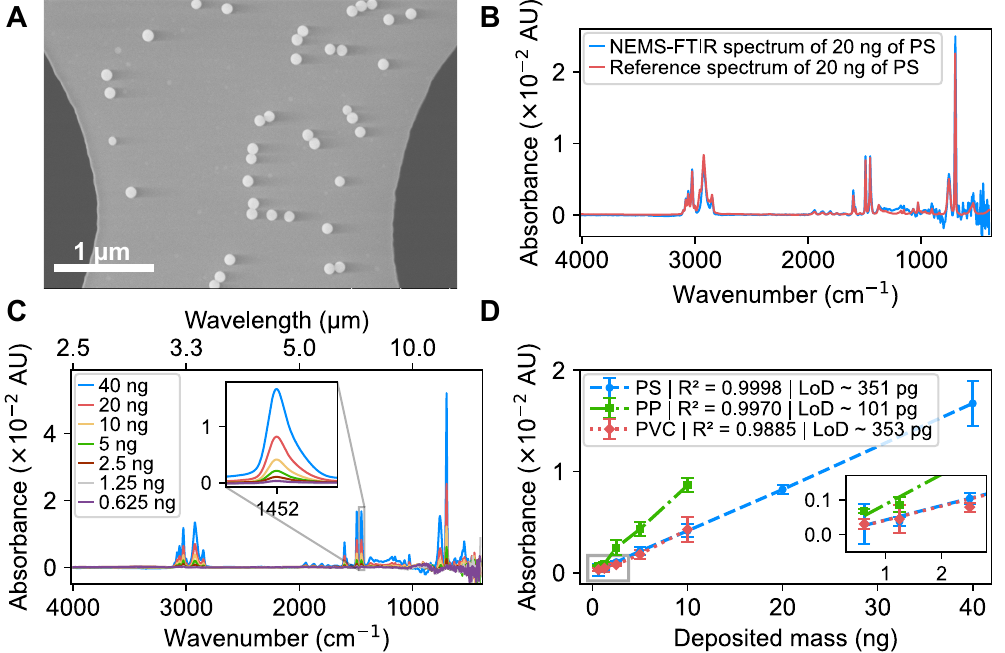}
\caption{\textbf{Characterization and quantification of model nanoplastics.} (\textbf{A}) SEM image of PS nanoparticles located on the perforated membrane area of the NEMS chip. (\textbf{B}) Absorbance spectrum of 20~ng PS measured \textit{via} NEMS-FTIR compared to a reference spectrum calculated with Eq.~(\ref{eq:absorbance-att coeff}) and (\ref{eq:dec lin att coeff}) from PS refractive index data.~\cite{Myers2018} (\textbf{C}) NEMS-FTIR spectra of varying mass loads of PS nanoparticles deposited on the NEMS chips. The inset highlights the 1452~cm$^{-1}$ peak, which was used to construct the calibration curve and determine the LoD. (\textbf{D}) Calibration curves for the 1452~$\text{cm}^{-1}$ PS peak, 1377~$\text{cm}^{-1}$ PP peak, and 1427~$\text{cm}^{-1}$ PVC peak, with the inset showing a zoomed-in view of the region corresponding to lower mass loads (error bars N=3, 95\% C.L.).}
\label{fig: Figure 2_PS}
\end{figure*}

Spectral data were obtained by depositing a 20~nL drop of various model nanoplastic concentrations on the NEMS chips (N=3 for each concentration). Fig.~\ref{fig: Figure 2_PS}B presents a comparison between the NEMS-FTIR spectrum of 20~ng of PS, expressed in absorbance, and a reference spectrum calculated from refractive index data \cite{Myers2018} for an equal sample mass.
Comparing the absorbance values of the two spectra enabled the calibration of the NEMS-FTIR method, determining $d_\text{IR}$, a model parameter representing the effective IR beam diameter illuminating the NEMS membrane. $d_\text{IR}$ is specific to the IR-light source. The extracted value, $d_\text{IR} =0.92 \pm 0.06$~mm, is smaller than the theoretical maximum of 1~mm set by the SiN membrane diameter, assuming a beam with a uniform intensity profile. This reduction indicates that the irradiance profile of the IR beam was highest at the center of the membrane, where the sample is located. The details of the spectral processing procedure are described in "Processing of NEMS-FTIR spectra" in the Materials and Methods section. 

In the NEMS-FTIR set-up, IR light from the source enters the vacuum chamber from below and passes through the resonator membrane before reaching the sample on its top surface (see inset in Fig.~\ref{fig: Figure 1}A). The resonator membrane exhibits a broad SiN absorption band with a maximum at 835~cm$^{-1}$ (see Fig.~S4). This feature can be exploited as an intrinsic internal standard to monitor chip-to-chip variability and measurement conditions. Because the SiN band absorbs approximately 20\% of the incident IR intensity at its maximum, sample vibrational modes that spectrally overlap with this region may, in principle, appear attenuated, as a fraction of the photon flux is already absorbed by the membrane. Such an effect, therefore, constitutes a potential measurement artifact that should be considered when interpreting spectra in this wavenumber range. Conversely, spectral overlaps could impact the internal standardization procedure. In the present measurements, however, no sample features overlap with the SiN absorption band (see Fig.~\ref{fig: Figure 2_PS}B) and no peak attenuation could be observed. As the spectral region around 835~cm$^{-1}$ is generally sparsely populated, this potential artifact is typically not an issue.
 
The NEMS-FTIR spectra of PS nanoplastic particles for each deposited mass, power-corrected, normalized using the SiN peak at 835~cm$^{-1}$, and converted into absorbance, are shown in Fig.~\ref{fig: Figure 2_PS}C. The spectra of PP and PVC nanoparticles are available in the Supporting Information in Figs.~S5 and S6, respectively. 

The IR peak positions in the recorded NEMS-FTIR spectra of PS nanoparticles are in agreement with data reported in the literature. \cite{Zajac2022PSpeaks, Wu2019PSpeaks, Smith2021PSandPPpeaks, NIST2025PSpeaks, Boke2022O-PTIR_PSPPPVCpeaks} Notably, PS is characterized by peaks at 753 and 700~$\text{cm}^{-1}$, as well as by a mono-substituted benzene ring, as evidenced by a series of weak overtone and combination bands between 2000 and 1650~$\text{cm}^{-1}$ (so-called 'benzene fingers'), often difficult to detect in the spectrum if the concentration of molecules containing the benzene ring is low. \cite{Smith2021PSandPPpeaks} However, these peaks are clearly visible in the NEMS-FTIR spectra of PS, even for the very low masses in the nanogram range. The exact peak positions are listed in Table~S1 in the Supporting Information.

The mass of PS nanoparticles present on the NEMS chips was estimated using the attenuation coefficient of PS and the absorptances calculated from the NEMS-FTIR spectra at the wavenumber of $\tilde\nu = 1452\text{ cm}^{-1}$. The mass estimation procedure is detailed in the "Mass estimation" in the Materials and Methods section. The calculation of the sample absorptance accounted for the spatial distribution of the particles, which predominantly formed a circular patch in the center of the perforation on the NEMS chips.
Fig.~S7 shows the correlation between the deposited (amount of deposited PS particles) and estimated (from the measured absorptances of PS particles) mass of nanoplastic particles.
The uncertainty in the sample mass deposited on the NEMS chip was estimated from the uncertainty in the tools used for sample preparation and for sample deposition. The largest uncertainty on the sample mass as estimated from the NEMS-FTIR spectra arises from the variability in the measured absorptance $\alpha_\text{S}(1452~\text{cm}^{-1})$. 
Exact values for the masses deposited on the NEMS chips and estimated masses as calculated from the NEMS-FTIR spectra, along with their standard deviations, can be found in Table~S2 in the Supporting Information. 

\begin{table*}
    \centering
    \captionsetup{width=\textwidth}
    \caption{\textbf{LoD calculation parameters.} Summary of the calculated LoDs for different nanoplastics, including the corresponding peak position (in wavenumbers), standard deviations of the blank ($\sigma$) at those respective peak positions (N = 9), and calibration slopes ($m$). LoDs were determined for $3\sigma/m$.}
    \label{tab:SD and slope}
    \resizebox{\linewidth}{!}{%
    \begin{tabular}{cccccc}
    \hline
    \\[-0.7em]
        \textbf{\parbox{2.4cm}{\centering Plastic Type}} & \textbf{\parbox{2.4cm}{\centering Peak position\\(cm$^{-1}$)}} & \textbf{\parbox{5.8cm}{\centering Standard deviation of the blank\\( $10^{-4}$ AU)}} & \textbf{\parbox{2.4cm}{\centering Slope\\($10^{-4}$ AU/ng)}} & \textbf{\parbox{2.4cm}{\centering LoD\\(pg)}} \\[0.7em]
        \hline
        \\[-0.9em]
        PS & 1452 & 0.49 &  4.17 & 351 \\
        PP & 1377 & 0.29 & 8.67 &  101 \\
        PVC & 1427 & 0.48 & 4.09 & 353 \\ [0.1em]
        \hline
    \end{tabular}%
    }
\end{table*}

The NEMS-FTIR spectra of PP and PVC nanoparticles also display peaks which are in agreement with prior studies using FTIR spectroscopy, \cite{karacan2011PPpeaks, Fang2012PPpeaks, Smith2021PSandPPpeaks, beltran1997PVCpeaks, Fawzi2025PVCpeaks, Boke2022O-PTIR_PSPPPVCpeaks} as well as with the measurements obtained using early IR spectrophotometers employing prisms. \cite{stromberg1958PVCpeaks} A detailed list of all characteristic peaks of PP and PVC nanoparticles, together with their spectral assignments and literature references, is provided in Table~S1 of the Supporting Information. NEMS-FTIR does not suffer from the red shift typically associated with measurements performed by ATR-FTIR,  as observed in previous studies. \cite{Tagg2017PPPVCPApeaks} In ATR-FTIR, slight shifts in some peak positions are observed due to the interaction of the evanescent wave with the sample surface, which depends on both the refractive index and wavelength. The wavelength-dependent penetration depth inherent to ATR measurements can also cause lower-wavenumber bands to appear more intense compared to transmission (or absorbance) spectra. \cite{bradley2018ATRpathlength,  mayerhofer2021ATR, grdadolnik2002ATR, miseo2025ATR, boulet2010ATR} This type of artifact is absent in NEMS-FTIR spectra.
NEMS-FTIR generates spectra comparable to those obtained with transmission-FTIR spectroscopy, allowing seamless integration with existing spectral libraries and chemometric tools for efficient and accurate analysis. Furthermore, since the nanoplastic particles are analyzed under high vacuum, the resulting spectra are free from spectral interferences from water molecules and carbon dioxide (CO$_2$).

Fig.~\ref{fig: Figure 2_PS}D shows the calibration curves for PS, PP, and PVC model nanoparticles. Several characteristic peaks were evaluated for each nanoplastic type to estimate the lowest limits of detection and highest coefficients of determination (R$^2$).
Calibration curves were constructed using the characteristic peaks at 1452~$\text{cm}^{-1}$ for PS (C-H bending vibrations; see inset in Fig.~\ref{fig: Figure 2_PS}C), 1377~$\text{cm}^{-1}$ for PP (methyl group, CH$_3$, symmetric bending; see inset in Fig.~S5), and 1427~$\text{cm}^{-1}$ for PVC (methylene group, CH$_2$, bending; see inset in Fig.~S6) and the results showed good linearity ($\text{R}^2>0.9885$). LoDs were calculated from the calibration curves using the relation $3\sigma/m$, where $\sigma$ is the standard deviation of method blanks ($\text{N}=9$) and $m$ is the slope of the corresponding calibration curve. The results are summarized in Table~\ref{tab:SD and slope}. PP exhibited the steepest calibration slope, resulting from the strongest IR band used for the calibration curve, and consequently had the lowest LoD. Detection limits are lower than those achieved by Py-GC/MS, with published values ranging from 1 to 10~ng \cite{Hasager2024PyGC-MS, Fischer2019PyGC-MS, li2024PyGC-MS} and are comparable to the performance of state-of-the-art TD-PTR-MS techniques with LOD as low as 340~pg. \cite{Materic2020TD-PTR-MS} 

The variations in detection limits for different nanoplastic particles did not correlate with nanoparticle size. In the measurements conducted on the chips with higher loadings, the observed nanoparticle agglomerations did not result in deviations in linearity.

While NEMS-FTIR can be used to detect very small masses of analytes, very dilute solutions will require pre-concentration. The largest droplet volume that can be drop cast on a NEMS chip is limited to approximately 500~nL. Considering an LOD of 351~pg for PS (see Table~\ref{tab:SD and slope}), the minimum detectable concentration of PS is 0.7~$\mu$g/mL for a single 500~nL drop of sample. However, in addition to usual pre-concentration techniques such as solid phase extraction, ultrafiltration, and diafiltration, the NEMS-FTIR chip allows new sample pre-concentration possibilities. Multiple droplets can be deposited consecutively on the NEMS chip with intermediate drying steps to increase the analyte mass deposited to the desired level. Sampling \textit{via} nebulization can also be used to concentrate larger sample volumes onto the NEMS chip. Based on a typical nebulizer flow rate of 20~$\mu$L/min, a sampling time of 1~hour, and an aerosol capture efficiency of approximately 50~\% for 100~nm particles, \cite{schmid2013real} we estimate that concentrations as low as 0.6~ng/mL could be measurable. For context, Li \textit{et al}.~\cite{Li2022tapwater} reported nanoplastic concentrations in tap water, with particle sizes ranging from 58 to 255~nm, varying between 1.67 and 2.08~ng/mL. When dealing with complex matrices containing high concentrations of organic material relative to the target analyte, matrix simplification or removal may be required to remove spectral interferences or avoid overloading the NEMS chip.

\subsection{Nanoplastic mixtures}
The SEM images in Fig.~\ref{fig: Figure 3_mixture}A, B provide the close-up views of the perforated area of the NEMS chip sampled with PS, PP, and PVC nanoparticles previously mixed in a 1:1:1 mass ratio. The particles highlighted in these images were identified based on their size, which corresponds to the known size distributions of the individual nanoplastics used: approximately 100~nm for PS, 54~nm for PP, and 262~nm for PVC.

The NEMS-FTIR spectrum of the mixture in Fig.~\ref{fig: Figure 3_mixture}C shows the characteristic peaks of each polymer component, highlighted by vertical lines. Despite a deposited mass of only 5~ng per component, the peaks corresponding to PS, PP, and PVC are clearly distinguishable. The exact peak positions of the PS, PP, and PVC nanoparticles can be found in the Supporting Information, in Table~S1. In environmental contexts, where mixtures of different polymer micro- and nanoparticles are commonly encountered, the broad spectral range of NEMS-FTIR increases the likelihood of detecting distinct peaks without interference, facilitating the identification of individual components when their spectral features are sufficiently resolved. The obtained spectra correspond to standard FTIR absorption spectra and can be deconvoluted \textit{via} chemometrics methods.

\begin{figure*}[t!]
\centering
\includegraphics[width=\textwidth]{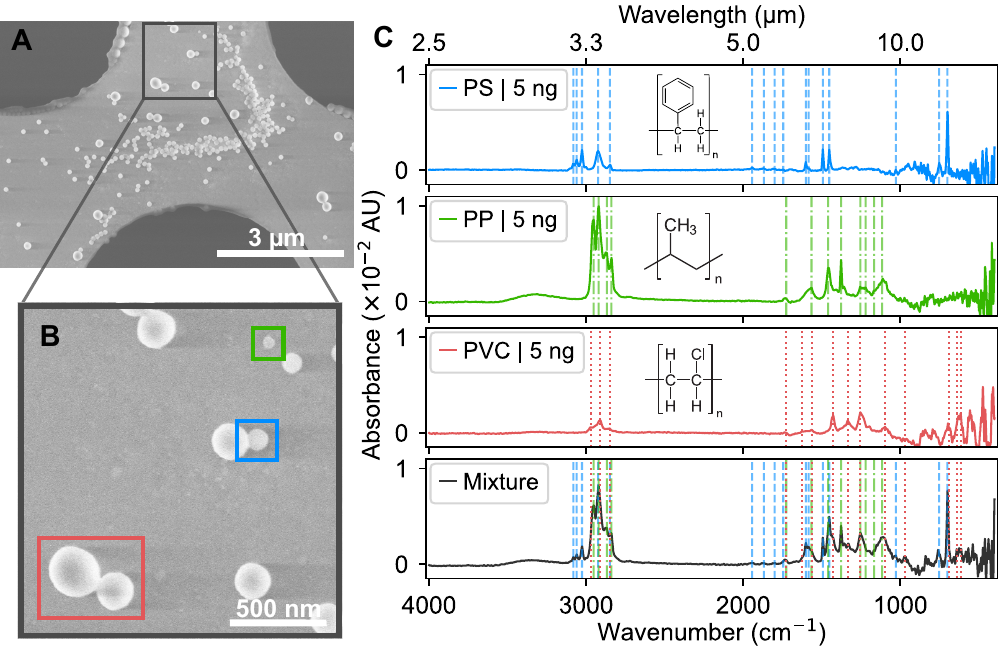}
\caption{\textbf{Characterization of nanoplastic mixture.} (\textbf{A}) SEM image of a mixture containing PS, PP, and PVC. (\textbf{B}) Magnified view of the membrane area highlighting PS (\diameter~100~nm, blue square), PP (\diameter~54~nm, green square), and PVC nanoparticles (\diameter~262~nm, red square). (\textbf{C}) Stacked NEMS-FTIR spectra of pure PS, PP, and PVC nanoparticles and their 1:1:1 mixture. Vertical lines mark the characteristic peaks of each polymer.}
\label{fig: Figure 3_mixture}
\end{figure*}

\subsection{Nylon teabag leachates}
Fig.~\ref{fig: Figure 4_tea}A shows the schematic of the experimental setup for the quantification of nylon teabag leachates. A single empty nylon teabag was immersed in 200~mL of water preheated at 95$\degree$C for 10 minutes. 

\begin{figure*}[t!]
\centering
\includegraphics[width=\textwidth]{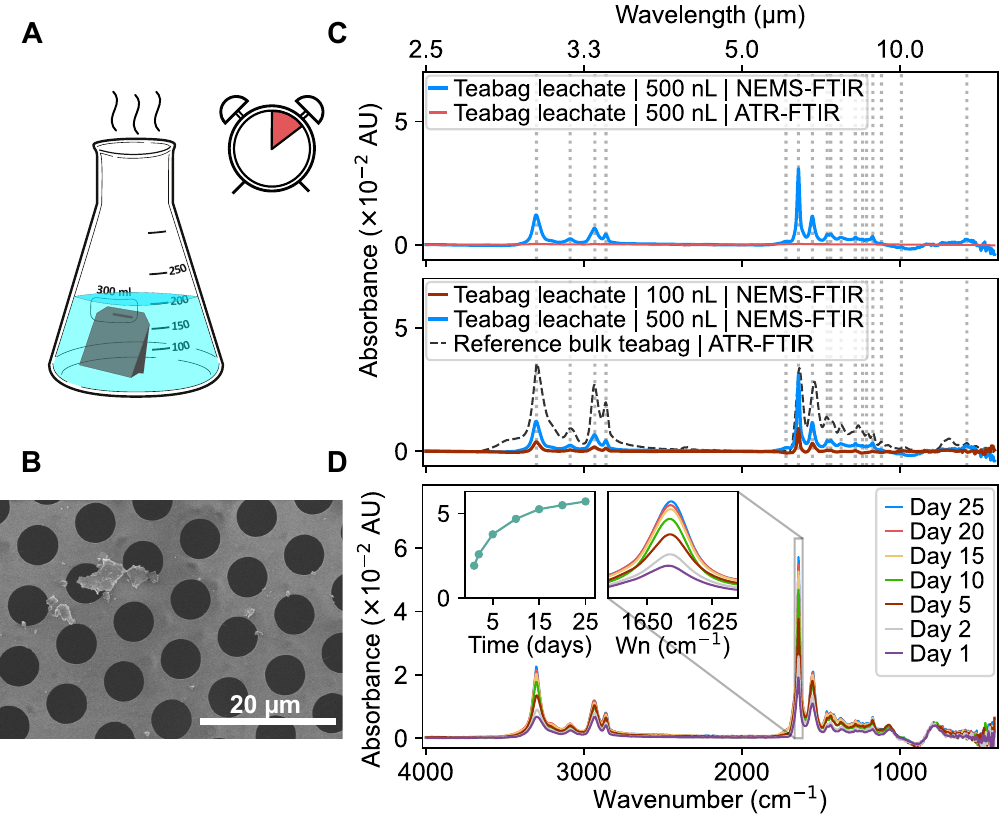}
\caption{\textbf{Characterization of leachates released from nylon teabags in water.} (\textbf{A}) Schematic illustration of the sample preparation process. (\textbf{B}) SEM image showing fragments released into water by a nylon teabag. (\textbf{C}) NEMS-FTIR spectra of nylon teabag leachates (100~nL, and 500~nL aliquots deposited onto NEMS chips without pre-concentration) compared to ATR-FTIR spectrum of 500~nL of teabag leachate and the ATR-FTIR spectrum of the bulk nylon teabag (Reference). Vertical lines indicate characteristic IR peaks associated with nylon-based PA. (\textbf{D}) NEMS-FTIR spectra of the leachates from nylon teabags subjected to accelerated weathering for 25 consecutive days, simulating one year of aging under adjusted conditions. The right inset graph highlights the characteristic nylon-based PA peak at 1642~cm$^{-1}$, corresponding to the amide I band. The left inset graph represents the dependence of signal intensity at 1642~cm$^{-1}$ on the number of days during which the nylon teabag was subjected to accelerated aging.}
\label{fig: Figure 4_tea}
\end{figure*}

Fig.~\ref{fig: Figure 4_tea}B shows micro- and nanoplastic fragments released from the nylon teabag into the water during brewing, which were deposited onto the NEMS chip using piezoelectric nanodroplet dispenser. Smaller fragments were also observed on the membranes after sampling the water from the aged teabags, where a different sample deposition method, based on aerosolization of the liquid, was used (see Fig.~S8).

Fig.~\ref{fig: Figure 4_tea}C shows the NEMS-FTIR spectra of nylon teabag leachates (N=3). 100~nL and 500~nL aliquots from the teabag leachates were deposited onto NEMS chips using the piezoelectric nanodroplet dispenser and the drop casting in combination with the pervaporation method, respectively, without prior concentration steps. A 500~nL aliquot of the teabag leachate was also analysed by ATR-FTIR (see Fig.~\ref{fig: Figure 4_tea}C). When compared with the NEMS-FTIR spectrum of the same 500~nL aliquot, the ATR-FTIR signal remains considerably weaker. Only faint indications of the most prominent nylon-based PA bands in the ATR-FTIR spectrum of the teabag leachate become visible upon amplification~($\times 20$), as shown in Fig.~S9 in the Supporting Information. In contrast, the reference ATR-FTIR spectrum of the bulk nylon teabag exhibits well-defined spectral features that correspond closely to those observed in the NEMS-FTIR spectra of the leachates, both for the 100~nL and the 500~nL aliquots (see Fig.~\ref{fig: Figure 4_tea}C). 

The NEMS-FTIR spectra collected from the teabag leachate allowed for chemical identification of the material from which the teabags were made and confirmation that this material was leaching into the tea water (see Fig.~\ref{fig: Figure 4_tea}C, D). Characteristic IR peaks, indicated by vertical lines in Fig.~\ref{fig: Figure 4_tea}C were observed, matching well with the reference ATR-FTIR spectrum of the original bulk teabag, as well as with the literature, \cite{Smith2023PApeaks, kang2022PApeaks, Tummino2023PApeaks, gonccalves2007PApeaksdegradation, Shurvell2001PApeaks, Zimudzi2018PApeakscarboxylicacid, OkambaDiogo2014PApeaksimidecarboxylicacid} confirming the presence of nylon-based PA, including the amide I (1642~cm$^{-1}$) and amide II (1553~cm$^{-1}$) bands. The amide I band primarily arises from C=O stretching vibrations in the amide group, while the amide II band results from a combination of N-H bending and C-N stretching vibrations, which are distinctive features of nylon-based materials \cite{kang2022PApeaks, Tummino2023PApeaks, Smith2023PApeaks}. In addition, several other characteristic nylon-related features were also observed in the NEMS-FTIR spectra. These include the CH$_2$ stretching vibrations at 2860~cm$^{-1}$ (symmetric) and 2930~cm$^{-1}$ (asymmetric), which are typical of the methylene groups in the nylon structure~\cite{kang2022PApeaks, Tummino2023PApeaks}. Distinctive peaks were detected at 3087~cm$^{-1}$, associated with N-H stretching overtones, and at 3300~cm$^{-1}$, corresponding to the N-H stretching vibrations of the amide group.~\cite{kang2022PApeaks, Tummino2023PApeaks, Smith2023PApeaks, gonccalves2007PApeaksdegradation, Tagg2017PPPVCPApeaks} A weak band at 1720 cm$^{-1}$, that is absent in the ATR-FTIR spectrum of the bulk teabag, appears in the NEMS-FTIR spectra of the leachate. Its presence is consistent with degradation-related functionalities such as imide \cite{gonccalves2007PApeaksdegradation} or carboxylic acid groups \cite{Zimudzi2018PApeakscarboxylicacid, OkambaDiogo2014PApeaksimidecarboxylicacid}, which can form during hot-water exposure and indicate the presence of hydrolysed chain ends or unbound low-molecular-weight oligomers. \cite{Kappenstein2018teabagoligomers, Heimrich2014teabagoligomers, Busse2020teabagoligomers} Exact positions and assignments for all observed peaks can be found in Table~S1 in the Supporting Information. The comparison with online spectral databases, such as Open Specy~\cite{cowger2021openspecy} and Spectragryph~\cite{menges2021spectragryph}, identified the released material as nylon-based PA.  

To compare NEMS-FTIR and ATR-FTIR, a correction factor was applied to the ATR-FTIR spectra. The uncorrected ATR-FTIR spectrum of the teabag is shown in Fig.~S10A, alongside the corrected spectrum to illustrate the effects of the applied automatic ATR correction procedure. As shown in Fig.~\ref{fig: Figure 4_tea}C, weak diamond phonon absorption bands originating from the ATR crystal \cite{miseo2025ATR} appear in the 2500–1800~cm$^{-1}$ region. A close-up of this spectrum is provided in Fig.~10B. These arise due to the refractive index contrast between the sample and diamond ($n = 2.42$~\cite{ralls1976introduction}). Since the refractive index of nylon is around 1.50–1.53,~\cite{shackelford2000materials} the difference ($\Delta n$) is sufficient to make the crystal phonon features visible in the spectrum. \cite{miseo2025ATR} Additionally, CO$_2$ absorption peaks are visible within the same region (see Fig.~10B), contributing to the observed spectral features. The variations in peak position and peak shapes, as well as the presence of additional diamond absorption bands associated with ATR-FTIR spectra, CO$_2$ peaks, and potential water vapor peaks, are all absent from the NEMS-FTIR spectra.

Differences in peak ratios were also observed between the ATR-FTIR spectrum of the bulk teabag and the NEMS-FTIR spectra of the 500~nL leachates (see Fig.~\ref{fig: Figure 4_tea}C). In NEMS-FTIR, the amide I peak at 1642 cm$^{-1}$ is enhanced compared to the ATR-FTIR spectrum. This likely reflects the differences between the samples, as the ATR-FTIR spectra were obtained from the intact bulk polymer, while NEMS-FTIR spectra were obtained from the leachate enriched in oligomeric nylon species released after exposure of the teabag to hot water. \cite{Busse2020teabagoligomers, Kappenstein2018teabagoligomers}

Nylon peaks appear in the NEMS-FTIR spectra of water samples in which individual empty nylon teabags were immersed and subjected to accelerated aging conditions. As aging progressed, the nylon teabags degraded further, releasing an increasing amount of nylon-based PA oligomers and small fragments, reflected in the peaks' increasing intensity as shown in the insets in Fig.~\ref{fig: Figure 4_tea}D. This observation is in line with previous studies suggesting that prolonged exposure to environmental aging conditions contributes to the release of polymeric particles into the surrounding medium. \cite{hernandez2023Accweathering, Montagna2023AccweatheringPA}

\begin{figure*}[t!]
\centering
\includegraphics[width=\textwidth]{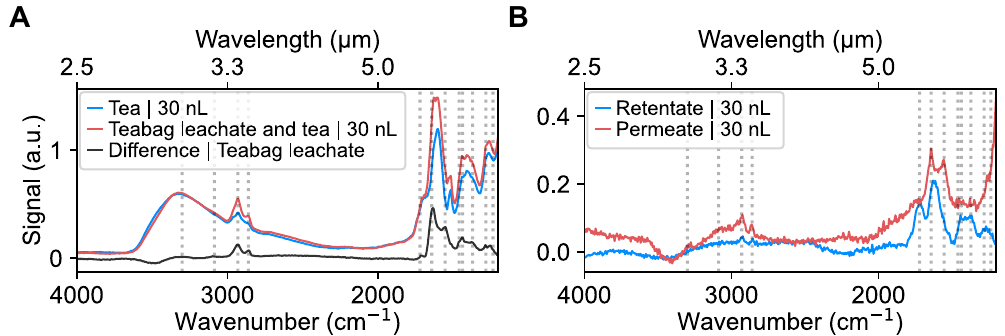}
\caption{\textbf{Characterization of leachates released from nylon teabags in a complex matrix.} 
(\textbf{A}) NEMS-FTIR spectra of brewed lemon balm tea with (red trace) and without (blue trace) the nylon teabag, together with the corresponding difference spectrum. Nylon-related peaks (vertical lines) remain distinguishable despite the complex matrix. (\textbf{B}) NEMS-FTIR spectra of the retentate, and permeate obtained after oxidation and ultrafiltration of the same samples. Nylon-based PA peak positions are marked with vertical lines. The permeate shows well-resolved nylon features, while the retentate spectrum is dominated by signals from the lemon balm tea matrix. All spectra are shown in arbitrary units.}
\label{fig: Figure 5_teabag in complex matrix}
\end{figure*}

The presence of characteristic nylon peaks could also be observed in brewed samples prepared with lemon balm tea leaves without digestion or ultrafiltration, as shown in Fig.~\ref{fig: Figure 5_teabag in complex matrix}A. The spectrum of this complex sample (red trace) shows a superposition of the nylon and lemon balm peaks. The nylon teabag spectrum (black trace) can be recovered by subtracting the signal from the pure lemon balm tea leaves (blue trace). Nylon-related peaks are marked with vertical dashed lines. The spectrum of pure lemon balm tea leaves exhibits characteristic features of plant-based polyphenolic extracts, including a dominant band at 1602~cm$^{-1}$ (C=C stetching), a carbonyl-related band near 1690~cm$^{-1}$, aromatic ring vibrations in the 1440-1400~cm$^{-1}$ region, a C-N stretching band near 1270~cm$^{-1}$, aliphatic CH/CH$_2$ stretching bands around 2900~cm$^{-1}$, and a broad O-H stretching feature near 3300~cm$^{-1}$ \cite{deOliveira2025lemonbalm, petenatti2014lemonbalm}.

The same lemon balm tea extract was brewed with a teabag, oxidized, and ultrafiltrated to remove the organic matrix \cite{foetisch2022teabags, hernandez2019teabags, Yousefi2024teabag, Yue2024teabag}. As shown in Fig.~\ref{fig: Figure 5_teabag in complex matrix}B, the NEMS-FTIR spectra of the resulting fractions (retentate and permeate) showed distinct differences. Although the retentate represents a concentrated fraction (reduced from 200~mL to 20~mL), its spectrum did not show clearly defined nylon peaks; instead, the observed features were largely dominated by signals originating from the lemon balm tea matrix. \cite{deOliveira2025lemonbalm, petenatti2014lemonbalm} This is likely due to residual fractions of the tea leaves remaining in the liquid, with particle sizes exceeding the ultrafiltration membrane’s nominal retaining size ($\approx$~1~nm).
In contrast, the NEMS-FTIR spectrum of the permeate displayed distinct, well-resolved nylon features despite approximately five-fold dilution due to the inclusion of the washing water (180~mL). These results indicate that smaller nylon-derived species, such as dissolved oligomers, passed through the ultrafiltration membrane. The spectrum shows characteristic peaks at 1553, 1642, 2860, 2930, and 3300~cm$^{-1}$, corresponding to typical vibrational modes of nylon-based PA and its methylene groups \cite{kang2022PApeaks, Tummino2023PApeaks, Smith2023PApeaks}.
The types and quantities of species released can also vary depending on how the nylon in the teabag was originally manufactured, including differences in polymerisation conditions or additives.

These observations are in agreement with LC-MS/MS analyses, which revealed the presence of nylon oligomers and monomers in teabag leachates.~\cite{Busse2020teabagoligomers, Kappenstein2018teabagoligomers} Comparable findings have been observed for PET teabags, with oligomers detected in the leachates. \cite{Tsochatzis2020teabagoligomers} Furthermore, research on other nylon materials has similarly reported the release of nylon oligomers into aqueous media. \cite{Heimrich2014teabagoligomers, Kappenstein2018teabagoligomers} These results indicate that the concentration of released plastics in complex samples may be underestimated when filtration and dialysis steps are used during the sample preparation if only the retentate is analysed using imaging techniques.

\section{CONCLUSIONS}

NEMS-FTIR has proven effective for rapidly identifying and quantifying nanoplastics in aqueous dispersions with picogram-level sensitivity. It was demonstrated that drop-cast or nebulized polymer nanoparticles could be identified from polydisperse mixtures and that the measured signals from deposited PS nanoparticles could be used to estimate the mass of PS present on the NEMS chip.

To demonstrate applicability to real-world samples, NEMS-FTIR was used to identify polymer particles and smaller oligomers released from nylon teabags during brewing. Leachates were directly analyzed on NEMS chips without prior sample preparation or preconcentration. We also show that for relatively simple aqueous matrices, such as tea infusions, identification of nylon leachates could be achieved without digestion, oxidation, ultrafiltration, or concentration steps. Established sample preparation methods can be used to remove concomitant species from more complex environmental samples, such as soil. Compared to ATR-FTIR, NEMS-FTIR offered significantly higher sensitivity, revealing distinct nylon-based PA spectral features from samples as small as 100~nL. ATR-FTIR, even with a five times larger volume, produced only a faint, barely distinguishable signal. Unlike ATR-FTIR, which requires additional spectral data corrections, and careful crystal selection, NEMS-FTIR simplifies data interpretation through direct absorption measurements. Measurements performed under high vacuum with a nitrogen-purged optical path minimize spectral interference from ambient CO$_2$ and water vapor, and the SiN material of the NEMS chip can be used as an intrinsic internal standard to account for chip-to-chip variability. Accelerated aging experiments on nylon teabags demonstrated the monitoring of the release of polymer particles and oligomers under prolonged environmental stress. It was shown that NEMS-FTIR spectra are compatible with standard transmission FTIR spectral databases, facilitating the reliable identification of unknown samples.

Taking advantage of the wide availability of FTIR instrumentation, NEMS-FTIR is ideally suited for routine monitoring and widespread adoption. Its non-destructive nature further allows seamless subsequent analysis by complementary techniques such as SEM, EDX, or O-PTIR for spatial sample characterization. The demonstrated performance and ease of integration with existing workflows highlight NEMS-FTIR as a promising tool for environmental monitoring and nanomaterials analysis in general.

\section{METHODS}

\subsection{Nanoplastic particles}
PS nanoparticles (\diameter 100 $\pm$ 10~nm, Prod. No. 43302), as a 10~\% w/v dispersion in water, were obtained from Sigma-Aldrich (MO, USA). PP (\diameter 54~nm, Prod. No. PP50) and PVC nanoparticles (\diameter 262 $\pm$ 30~nm, Prod. No. PVC250), both as 1~\% w/v dispersion in water, were obtained from Lab261 (CA, USA). 

\subsubsection{Preparation of single and mixed nanoplastic dispersions}\label{subsubsec:pureandmixed}
All equipment used was first cleaned with acetone (HPLC Plus grade, Prod. No. 650501, Sigma-Aldrich) and isopropyl alcohol (electronic grade, Prod. No. 733458, Sigma-Aldrich), then rinsed five times with distilled deionized water (DDW, 18~M$\Omega$cm; Millipore, MA, USA) followed by UHPLC-MS grade water (Prod. No. W81, Thermo Fisher Scientific, MA, USA). Finally, all items were allowed to dry under a laminar flow hood (Envirco, NC, USA) to prevent contamination. All samples were prepared in a fume hood (Secuflow, Waldner Holding, Germany) while wearing nitrile gloves (Prod. No. 0439, Helmut Feldtmann GmbH, Germany). Clean nitrile gloves were also used during all experimental procedures.

Serial dilutions and mixture preparation were carried out using UHPLC-MS grade water. A PIPETMAN P1000 micropipette, with matching DIAMOND D1000 TIPACK tips (Prod. No. F144059M and F171500, respectively, Gilson Incorporated, WI, USA), was used for the preparation.  All dilutions were stored in hydrolytic class 3 soda-lime glass vials with snap-on lids (Prod. No. LC84.1 and LC87.1, respectively, Carl Roth GmbH, Germany).

All stock and diluted dispersions were first vortexed for 30 seconds using a Mini Vortexer (Heathrow Scientific, IL, USA) to ensure consistent distribution before proceeding with the successive dilution. PS nanoparticle dispersions were prepared at concentrations of 31.25-2000~$\mu$g/mL, while PP and PVC dispersions ranged from 31.25 to 500~$\mu$g/mL. The mixed dispersion of all three types of polymer nanoparticles was prepared in a 1:1:1 mass ratio (PS:PP:PVC). Table~\ref{tab:NPs_concentrations_masses} summarizes all final concentrations used for each single and mixed sample. Blank samples, consisting of the same UHPLC-MS-grade water used for dilution procedures, underwent the same preparation process as the experimental samples.

\begin{table*}[!b]
\centering
\captionsetup{width=\textwidth}
\caption{\textbf{Characterization of nanoplastic dispersions.} Concentrations of dispersions (A) and total deposited masses of nanoplastic particles (B), differentiating between single- and three-component dispersions. Empty cells indicate concentrations not used in the experiment for the corresponding dispersion.}
\label{tab:NPs_concentrations_masses}
\normalsize
\begin{tabularx}{\textwidth}{l*{7}{>{\centering\arraybackslash}X}}
\hline
\multirow{3}{*}{\textbf{Nanoplastics type}} & \multicolumn{7}{c}{\textbf{Concentration ($\mu$g/mL) (A)}} \\
\cline{2-8}
 & 2000 & 1000 & 500 & 250 & 125 & 62.5 & 31.25  \\
\cline{2-8}
 & \multicolumn{7}{c}{\textbf{Total deposited mass (ng) (B)}} \\
\hline
PS & 40 & 20 & 10 & 5 & 2.5 & 1.25 & 0.625  \\
PP &  &  & 10 & 5 & 2.5 & 1.25 & 0.625  \\
PVC &  &  & 10 & 5 & 2.5 & 1.25 & 0.625   \\
PS:PP:PVC (1:1:1) &  &  &  & 5 &  &  &   \\
\hline
\end{tabularx}
\normalsize
\end{table*}

\subsection{Teabags}
\subsubsection{Standard tea brewing procedure}
\textbf{\textit{Teabags in water}}\\
The teabags used in the experiment were purchased online (Temu.com, China), and the procedure for releasing plastic particles from the teabags was adapted from the study by Hernandez \textit{et al}.~\cite{hernandez2019teabags}
To minimize potential contamination, the teabags, made of nylon according to the manufacturer, were purchased empty, eliminating the need for additional cutting or removing the tea leaves. These empty teabags were first used to establish a controlled reference system, before introducing the more complex matrix containing tea leaves. The thin cotton string that attaches a tiny label to the teabag was carefully removed with scissors. The teabags were then rinsed five times with UHPLC-MS grade water to avoid impurities or loose plastic particles and allowed to dry overnight in a laminar flow hood.

To simulate a typical tea brewing process, 200~mL of UHPLC-MS grade water was poured into an Erlenmeyer flask (Duran, DWK Life Sciences GmbH, Germany), which had been pre-cleaned following the same procedure described in the "Preparation of single and mixed nanoplastic dispersions" section. Once the water reached its boiling point, the flask was removed from the hot plate and a single teabag was immersed. After 10~minutes, the teabag was carefully removed using clean metal tweezers, and the water was allowed to cool overnight. The flask was covered with aluminum foil to prevent contamination during the cooling period. A blank sample was prepared using the same procedure with the same UHPLC-MS grade water but without the teabag.\\
\textbf{\textit{Teabags and tea}}\\
To evaluate the performance of NEMS-FTIR in a more complex matrix containing natural organic matter, fresh lemon balm tea leaves were brewed together with a nylon teabag. The procedure was adapted from Foetisch et al.~\cite{foetisch2022teabags}

First, 200~mL of UHPLC-MS grade water was boiled, after which approximately 1~g of fresh lemon balm leaves was placed in a metal tea sieve (pore size not specified) and immersed in the water at 95$\degree$C for 5~min. The resulting infusion was divided into two 25~mL aliquots, each of which was reheated to boiling.

A single nylon teabag was added to one of the aliquots and steeped for 5~minutes before being removed with metal tweezers, mimicking a brewing scenario in which both leaves and teabag were present. The second aliquot, containing only tea leaves, served as a control matrix representing organic material derived from the leaves without the contribution of the teabag. After steeping, both aliquots were allowed to cool for several hours while covered with aluminum foil to avoid contamination. Both resulting infusions were then analyzed by NEMS-FTIR post-brewing, and before any organic digestion and ultrafiltration process.

Organic digestion was performed on both samples by six sequential additions of hydrogen peroxide (H$_2$O$_2$; 30\%, Prod. No. CP26.1, Carl Roth GmbH, Germany). The amount of H$_2$O$_2$ added was calculated to maintain an overall concentration of approximately 5\%, assuming complete consumption of the previously added H$_2$O$_2$.

After digestion, both samples were subjected to ultrafiltration and washing using an Amicon Stirred Cell (200~mL, Prod. No. UFSC20001, Sigma-Aldrich) fitted with a 3~kDa Ultracel membrane (dia. 63.5~mm, Prod. No. PLBC06210, Sigma-Aldrich). The membrane is characterized by a nominal retaining size of approximately 1~nm.~\cite{SigmaAldrich2025ultrafiltration} Nitrogen gas at 2.3~mbar was used to push the samples through the membranes. The stirred cell reservoir was filled with UHPLC-MS grade water up to 200~mL and filtered until approximately 20~mL remained. Processing approximately 180~mL of liquid required about 2~h. Both the retentate and the permeate were collected for NEMS-FTIR analysis.

\subsubsection{Accelerated aging of teabags}
The accelerated aging experiment was carried out in a SimTech Feutron Double Climate Chamber (73 x 77 x 102~cm, Feutron Klimasimulation GmbH, Germany) equipped with two Ultra-Vitalux lamps (300~W, Prod. No. 4008321543929, Osram, Germany) placed 30~cm apart (see Fig.~S11 in the Supporting Information). These lamps emitted broad-spectrum UV light (UVA: 315-400~nm, UVB: 280-315~nm) with an irradiance of approximately 59.1~W/m$^2$, simulating natural solar radiation to induce degradation.

A metal plate, 32~cm below the lamps, held hydrolytic class 1 soda-lime glass vials with screw caps (Prod. No. LC92.1 and LE03.1, respectively, Carl Roth GmbH, Germany) containing the teabag samples. The vials allowed good UV transmission, particularly in the UVA range, while minimizing light absorption.

The chamber was operated according to a cyclic schedule: 8 hours of UV exposure at 60\degree C, followed by 4 hours in the dark at 25\degree C, mimicking natural day and night temperature fluctuations. This regimen, based on ISO 4892-3~\cite{iso4892-3} and ASTM G154~\cite{astmG154} standards, lasted 25 days, which is equivalent to approximately one year of natural aging. \cite{philip2018AccWeathering} Seven nylon teabags, each in one glass vial with 30~mL of UHPLC-MS water, were artificially aged, and individual water samples were collected after 1, 2, 5, 10, 15, 20, and 25 day(s) to monitor particle release and degradation. Simultaneously, blanks (water without teabags) were collected with each respective teabag sample.

\subsection{NEMS chips}
As shown in Fig.~\ref{fig: Figure 1}B, NEMS chips, or resonators, play a dual role as both sensing devices and sample holders, making them the core of NEMS-FTIR technology. The wafers used in this study to fabricate NEMS chips are made from silicon (Si) and SiN, with thicknesses of 380~$\mu$m and 50~nm, respectively, and an intrinsic tensile stress of approximately 50~MPa for the SiN. After the final manufacturing step, the KOH etching, a 1x1~mm$^2$ SiN membrane is released. This membrane has a perforated area with a diameter of approximately 600~$\mu$m, consisting of 6~$\mu$m holes spaced 3~$\mu$m apart. It acts as the platform on which the sample is deposited for analysis. Two gold electrodes, each 10~$\mu$m wide, are positioned across the chip to enable electrical transduction, allowing mechanical vibrations to be converted into electrical signals, as demonstrated by Be\v{s}i\'c \textit{et al}.~\cite{ besic2024NEMS-IR} and Luhmann \textit{et al}.~\cite{Nik2023thermaldesorption} After fabrication, the disposable NEMS chips are stored in individual closed plastic containers to prevent cross-contamination and non-specific adsorption from the environment.

\subsection{Sampling methods}
\subsubsection{Drop casting methods}
Various drop casting strategies may be applied depending on the specific requirements of sample deposition. In this work, two different drop casting approaches were employed to deposit samples onto NEMS chips, selected based on the target deposition volume.

In the nanodroplet dispensing approach, a piezoelectric nanodroplet dispenser (PIPEJET nanoDispenser, BioFluidix-Hamilton Freiburg, Germany), shown in Fig.~S12A, was used to deposit the dispersions on the membrane of the NEMS chips. Polyimide capillaries with an inner diameter of 200~$\mu$m (Prod. No. PJ-20010) were used to deliver individual droplets of 20~nL. A 1~cm length of Tygon tubing (Prod. No. VM-20053-1) was attached to the capillary to facilitate sample loading. Liquid samples were inserted into the tubing using a micropipette (PIPETMAN P1000, Prod. No. F144059M, Gilson Incorporated, WI, USA) with matching micropipette tips (DIAMOND D1000 TIPACK tips, Prod. No. F171500, Gilson Incorporated, WI, USA). The stroke velocity was set to 90~$\mu$m/ms during nanodroplet dispensing. 
For each concentration of the dilution series, droplets of 20~nL were deposited onto the center of the individual NEMS chips. The total deposited mass of each sample is listed in Table~\ref{tab:NPs_concentrations_masses}B. Each deposition was performed in triplicate, with three NEMS chips prepared for each specific polymer mass. Nine additional NEMS chips were sampled with blank samples to provide a reliable baseline for subtraction.
For the mixture of nanoplastic particles, a single replicate was prepared by depositing one 20~nL drop of the pre-mixed dispersion onto a NEMS chip. Blank samples were deposited in parallel onto three additional NEMS chips.
For the analysis of the teabag leachates in water, three NEMS chips were sampled with leachate samples and three with blank samples using the piezoelectric nanodroplet dispenser. Each NEMS chip was sequentially loaded with ten drops (10~nL each) of liquid sample, with a two-minute delay between drops, totaling 100~nL per chip.

In the approach combining the drop casting and the pervaporation method, three NEMS chips were prepared using the Drop Casting Accessory (Invisible-Light Labs GmbH, Austria), a micropipette (PIPETMAN P2, Prod. No. F144054M, Gilson Incorporated, WI, USA), and corresponding micropipette tips (DIAMOND DL10 TIPACK, Prod. No. F171200, Gilson Incorporated, WI, USA). A single 500~nL droplet of the teabag leachate was deposited onto each chip. The chips were then left to dry inside the Drop casting Accessory, as shown in Fig.~S12B, in an environment with a controlled humidity gradient, which, due to the presence of perforations in the chip, facilitates droplet drying within the perforated area \textit{via} pervaporation. The approximate drying time for a 500~nL droplet under these conditions was 30~minutes. In parallel, three additional NEMS chips were sampled using blank samples.

For the analysis of the complex aqueous matrices, single replicate samples were prepared using the piezoelectric nanodroplet dispenser by depositing three drops of 10~nL each, totaling 30~nL, onto each NEMS chip. The liquid samples consisted of unprocessed tea leaves and teabag extracts prior to any sample preparation steps as well as after organic digestion and ultrafiltration. Corresponding single replicate blank samples, prepared from tea leaves extracts without teabag material, were deposited in the same manner.

\subsubsection{Aerosolization method for particle deposition}
A system incorporating the Portable Aerosol Generation System (PAGS, Handix Scientific Inc., CO, USA) was employed to sample particles released from the nylon teabags subjected to accelerated weathering cycles. This setup enabled consistent aerosolization and deposition of particles onto the NEMS chips facilitated by the holes in the central region of the membrane. Sampling was conducted at regular intervals (day~1, day~2, day~5, day~10, day~15, day~20, and day~25) to monitor the temporal evolution of particle release. The NEMS chip was placed in a holder (Aerosol Flow Adapter, Invisible-Light Labs GmbH, Austria), which was then positioned in an eight-channel filter sampler (FILT, Brechtel Manufacturing Inc., CA, USA) equipped with a built-in pump. A schematic of the setup is presented in Fig.~S12C. After nebulization, the aerosolized droplets containing the extracted particles are drawn through the central perforation of the NEMS chip. These small openings constrain and accelerate the airflow, increasing particle momentum. Particles with sufficient inertia no longer follow the bending streamlines around the membrane but instead depart from the flow and impact directly onto the perforated area.\cite{schmid2013real, Kurek2017pharmaceuticalcompounds, Nik2023thermaldesorption} Due to the inertial impaction mechanism, the deposited sample covers the entire perforated area of the membrane, as shown in Fig.~S1B. The pump was set to a flow rate of 0.5~L/min for 5~minutes for each NEMS chip.

\subsection{Experimental setup}
\subsubsection{SEM imaging}
SEM (Hitachi SU8030, Hitachi, Japan) was employed to confirm the sizes of the particles deposited on the NEMS chips and to analyze the leachates from the plastic teabags. For imaging, the sampled NEMS chips were mounted on an aluminum sample holder. Secondary electron (SE) detection in the upper lens mode was utilized to capture high-resolution surface details. The microscope operated at an accelerating voltage of 2~kV with an emission current of 3900~nA. NEMS chips were examined at various magnifications to assess the particle dimensions and the structural characteristics of the deposited material.

\subsubsection{NEMS-FTIR spectroscopy}
A nanoelectromechanical IR analyzer (EMILIE$^{\text{TM}}$, Invisible-Light Labs GmbH, Austria) in conjunction with an FTIR spectrometer (Vertex 70, Bruker Optics, MA, USA), as shown in Fig.~\ref{fig: Figure 1}A, was used to collect all the NEMS-FTIR spectra and confirm the chemical identity of different types of nanoplastic particles, their mixture, and teabag leachates.

During the measurement process, sampled NEMS chips were placed in the vacuum chamber of the nanoelectromechanical IR analyzer at 10\textsuperscript{-5}~mbar and the temperature of the chips was consistently regulated to 25$\degree$C using the integrated thermoelectric cooling (TEC) controller and Peltier element. To minimize atmospheric interferences, the optical path of the EMILIE$^{\text{TM}}$ and FTIR spectrometer were purged with dry nitrogen throughout the measurements. A resolution of 4~cm$^{-1}$, with a stabilization delay of 30~ms, 200 co-additions, and an aperture of 6~mm, were used for the FTIR parameters. The spectral range extended from 4000 to 400~cm$^{-1}$.
The spectrometer aperture was chosen to be sufficiently large, such that the IR beam spot is always significantly larger than the membrane area. For each individual chip, the IR beam was fine-aligned to ensure maximal signal amplitude.

\subsubsection{ATR-FTIR spectroscopy}
ATR-FTIR spectroscopy was used alongside the NEMS-FTIR to analyze the teabag leachates as a reference IR technique. The teabag material itself was also examined using ATR-FTIR to confirm it was the source of the detected components in the leachates. A Platinum ATR accessory with a diamond crystal (A225/Q, Bruker Optics, MA, USA) was used for all ATR measurements, performed on the same Bruker Vertex 70 spectrometer utilized for the NEMS-FTIR experiments. All ATR-FTIR measurements were performed under nitrogen-purged conditions to ensure consistency with the NEMS-FTIR measurement environment.

To match the NEMS-FTIR experimental conditions, an aliquot of 500~nL of teabag leachate was applied to the isopropanol-cleaned ATR diamond crystal. The droplet was dispensed using the same micropipette and tip combination as that used for sampling the NEMS chips with the same teabag leachates (as described in the "Drop casting methods" section). Once the droplet had evaporated, the ATR measurement was performed by lowering the pressure clamp onto the remaining residue. Additionally, 500~nL of the corresponding blank sample was measured in the same manner.

A section of the teabag material used in the brewing experiments was also analyzed. The cleaned ATR crystal was covered with the teabag fragment, and the pressure clamp was lowered to ensure a good contact between the sample and the crystal. 

Each measurement consisted of a background spectrum recorded with the empty crystal under pressure, followed by the sample spectrum (blank, leachate, or teabag). All spectra were acquired with a resolution of 4~cm$^{-1}$, over the spectral range of 4000-400~cm$^{-1}$, using 32 scans, a scanner velocity of 2.5~kHz, and an aperture of 6~mm. Atmospheric compensation was enabled during all measurements to minimize the influence of the water vapor and CO$_2$ absorption.

The ATR-FTIR spectra were processed using the "Extended ATR correction" (standard mode) and "Baseline correction" features in the OPUS software to correct for the slight distortions in peak intensity, shape, and position inherent to ATR-FTIR measurements. \cite{bradley2018ATRpathlength,  mayerhofer2021ATR, grdadolnik2002ATR, miseo2025ATR, boulet2010ATR} The spectra were also smoothed using Savitzky–Golay filtering (window length 20, polyorder 2).

\subsection{Processing of NEMS-FTIR spectra}

All raw NEMS-FTIR spectra were smoothed using a Savitzky-Golay filter with a window length of 20 and a polynomial order of 2. The spectra of the blanks and analytes were then divided by the recorded NEMS-FTIR spectrum of the FTIR light source to account for the wavelength-dependent intensity of the light source. The NEMS-FTIR spectrum of the light source was recorded using a NEMS chip coated with a 5~nm ultra-thin Pt film (EMILIE$^{\text{TM}}$ LIGHT chip, Invisible-Light Labs GmbH, Austria), which acts as a broadband absorber. \cite{piller2022thermal} Following this correction, the spectra were scaled by a normalization factor to set the SiN peak at 835~cm$^{-1}$ to a value of 1 $(R_\text{SiN} = 1)$.

Spectral calibration was carried out using the intrinsic SiN absorption peak at 835~cm$^{-1}$ as a reference. This calibration was based on transmission measurements through 12 clean, perforated, $2\times2$~mm$^2$, 50~nm thick low-stress SiN NEMS membranes.
Fig.~S4 in the Supporting Information shows the average transmittance and corresponding absorptance of the NEMS chip's SiN material measured using the internal detector of the FTIR spectrometer (2~mm aperture, 32 scans, and a resolution of 2~cm$^{-1}$). Despite the different sizes of the membranes ($1\times1$~mm$^2$ and $2\times2$~mm$^2$) and perforation areas (600~$\mu$m and 1200~$\mu$m, respectively), the ratio of perforated to non-perforated membrane material remains consistent across both membrane types, making the calibration based on the SiN absorption peak possible.

The absorptance for a specific wavenumber $\tilde{\nu}$, $\alpha(\tilde\nu)$, was calculated from the transmittance $T(\tilde\nu)$ as $\alpha(\tilde\nu) = 1- T(\tilde\nu)$, assuming insignificant scattering, yielding $\alpha_{\text{SiN}}(835 \text{ cm}^{-1}) = 0.21 \pm 0.01$. 
The NEMS-FTIR response of the bare SiN chip, $R_\text{SiN}$, was then related to absorptance through a calibration factor $\beta$:
\begin{equation}
\beta = \frac{\alpha_{\text{SiN}}(835 \text{ cm}^{-1})}{R_{\text{SiN}}(835 \text{ cm}^{-1})}.
\label{eq:beta}
\end{equation}

Depositing the nanoplastics on the membrane using the piezoelectric nanodroplet dispenser created a circular sample spot (see Fig.~S2 in Supporting Information). The nanoplastics occupied the area $\Sigma_\text{S}$, which is smaller than the area of the IR beam illuminating the chip, $\Sigma_{\text{IR}}$. Furthermore, the increased photothermal response, if the sample is concentrated in the membrane center instead of being evenly distributed over the entire membrane, was corrected by $\gamma$. \cite{schmid2023fundamentals} 
The sample absorptance $\alpha_\text{S}(\tilde{\nu})$ was calculated from the measured  NEMS-FTIR signal of the sample, $R_\text{S}$, as:
\begin{equation}
\alpha_\text{S}(\tilde{\nu}) = \beta R_{\text{S}}(\tilde{\nu}) \frac{\Sigma_{\text{{IR}}}}{\Sigma_{\text{S}}} \frac{1}{\gamma}.
\label{eq:sample-absorptance}
\end{equation}

$\Sigma_{\text{IR}}$ was calculated based on the experimentally determined effective IR beam diameter, $d_\text{IR}$. For each deposited mass of the PS nanoparticles, $d_\text{IR}$ was iteratively adjusted to optimize the fit between the measured NEMS-FTIR spectrum and a reference spectrum generated from the known refractive index of PS. \cite{Myers2018} The optimization focused particularly on matching the intensity and shape of the characteristic absorption peak at 1452~cm$^{-1}$. This method provided an estimate of the effective beam diameter, which represents the actual size of the IR spot that effectively illuminates the SiN membrane. $d_\text{IR}$ was found to be $0.92 \pm 0.06$~mm, which is close to the ideal value corresponding to the SiN membrane with a lateral dimension of 1~mm, and the total area illuminated by the IR beam was calculated as $\Sigma_{\text{IR}} = \pi (d_\text{IR}/2)^2 - \Sigma_\text{P}$, with the area of the perforation $\Sigma_\text{P}=0.108~\text{mm}^2$.

$\Sigma_{\text{S}}$ corresponds to the diameter of the central nanoplastics area on the membrane with a diameter of $d_\text{S}=504\pm46$~$\mu$m. These measurements were obtained by analyzing six separate NEMS chips, each imaged under a microscope. $d_\text{S}$ was then quantified from these images using ImageJ software. The total sample area is then given by $\Sigma_{\text{S}} = \eta \pi (d_\text{S}/2)^2$, with the fill factor of the perforation $\eta=0.62$. 

$\gamma$ was calculated by finite element method (FEM) simulations by Kanellopulos \textit{et al}.~\cite{kanellopulos2025responsivity} The results for a square low-stress SiN membrane with a side length of 1~mm are given in Fig.~S13 in the Supporting Information, yielding  $\gamma = 1.68 \pm 0.05$ for the given sample spot size. Thus, the response is 1.68 times greater than it would be for a sample uniformly distributed across the entire square membrane.
The nanoplastics spots on the membrane formed a coffee ring, as shown in Fig.~S2 in the Supporting Information. Up to 90\% of particles typically accumulate at the droplet perimeter in the coffee ring, as shown for nanometer-sized PS dispersions. \cite{li2016coffeering, rey2022coffeering} FEM simulations have shown that the formation of a coffee ring has no significant effect on the NEMS-FTIR response (see Fig.~S14), with an estimated uncertainty in the responsivity of < 3\%.

Finally, the blank-corrected nanoplastic absorbance can be calculated from
\begin{equation}
    A(\tilde{\nu}) = -\log_{10}[1-\alpha_{\text{S}}(\tilde{\nu})] + \log_{10}[1-\alpha_{\text{B}}(\tilde{\nu})],
    \label{eq:absorbance-absorptance}
\end{equation}
with the absorptance of the blank $\alpha_{\text{B}}(\tilde\nu)$.

For samples containing a complex matrix, a modified subtraction procedure was required to account for differences in the tea-leaf background between the blank and analyte spectra. Although both spectra were prepared from the same initial tea infusion by splitting it into two aliquots, where one was used as the blank and the other was enriched with an empty nylon teabag, the background contributions from the leaf extract were not identical. This likely stems from the heterogeneous nature of the organic material. Despite using fresh leaves to minimize fragmentation, small particulate or colloidal components may have been unevenly distributed between the aliquots. To eliminate the matrix-dependent spectral offset,  spectra were normalized over a region between 3700 and 2500~cm$^{-1}$ and subtracted. As a result, the processed spectra are presented in arbitrary units.

\subsection{Mass estimation}
This framework aims to estimate the actual mass of the nanoplastic on the NEMS chip responsible for the NEMS-FTIR signal. Following the approach of Dudani and Takahama,~\cite{satoshi2022quantitativeevaluation} it was assumed that particle scattering is negligible compared to absorption; thus, correction factors related to scattering are not required. Particles are much smaller than the wavelength of IR radiation used in this study, placing them in the Rayleigh regime, where absorption dominates IR attenuation. In this case, the absorbance $A(\tilde\nu)$ of the deposited particles can be expressed as a function of the linear attenuation coefficient $\mu_\text{10}(\tilde\nu)$ :
\begin{equation}
    A(\tilde\nu) = \mu_\text{10}(\tilde\nu) h_\text{eff}=\mu_{\text{10}}(\tilde\nu)\dfrac{m}{\rho \Sigma_\text{S}},
    \label{eq:absorbance-att coeff}
\end{equation}
where the effective film thickness $h_\text{eff}$ is expressed as a function of the analyte mass $m$, the nanoparticle material density $\rho$, and the sample area $\Sigma_\text{S}$.

In NEMS-FTIR, the measured absorptance values are small $(\alpha_\text{S}(\tilde\nu)\ll 1)$ and the absorbance (\ref{eq:absorbance-absorptance}) can be approximated using a first-order Taylor series expansion:
\begin{equation}
    A(\tilde{\nu}) \approx \frac{\alpha_\text{S}(\tilde{\nu})-\alpha_\text{B}(\tilde{\nu})}{\ln(10)}.
    \label{eq:absorbance-Taylor approx}
\end{equation}
Combining Eqs. (\ref{eq:sample-absorptance}),  (\ref{eq:absorbance-att coeff}), (\ref{eq:absorbance-Taylor approx}), and solving for the analyte mass gives: 
\begin{equation}
    m = \beta \frac{R_\text{S}(\tilde\nu)-R_\text{B}(\tilde\nu)}{\mu_\text{10}(\tilde{\nu})}\frac{\rho}{\ln(10)}\frac{\Sigma_\text{IR}}{\gamma},
    \label{eq:analyte mass}
\end{equation}
where $R_\text{B}(\tilde\nu)$ is the NEMS-FTIR response of the averaged blank spectra, and $\tilde{\nu}$ denotes the wavenumber corresponding to a characteristic vibrational mode of the sample.

The attenuation coefficients of the substance are determined from its complex refractive index, defined as $\tilde{n} = n + ik$, where $n$ and $k$ are the real and imaginary components, respectively. For a sparsely distributed particle layer, the decadic linear attenuation coefficient is given by:

\begin{equation}
\mu_{\text{10}}(\tilde \nu) = \dfrac{6\pi \tilde\nu}{\ln{(10)}}\Im{\dfrac{\tilde n^2(\tilde\nu)-1}{\tilde n^2(\tilde\nu)+2}}.
\label{eq:dec lin att coeff}
\end{equation}

A detailed derivation of Eq.~(\ref{eq:dec lin att coeff}) can be found in the Supporting Information, section A.

\section{Supporting Information}
\begin{footnotesize}
\noindent\textbf{This file includes:}\\
Derivation of the decadic linear attenuation coefficient for sparse particle film, sample distribution on the membrane of the NEMS chip after aerosol-based deposition and drop casting in conjunction with pervaporation, sample distribution on the membrane of NEMS chip after using the nanodroplet dispenser to deposit the nanoparticles, SEM images of polypropylene and polyvinyl chloride nanoparticles sitting on the perforated area of the membrane, IR spectral features of silicon nitride membranes, NEMS-FTIR spectra of polypropylene nanoparticles, NEMS-FTIR spectra of polyvinyl chloride nanoparticles, correlation between the deposited (amount of deposited PS particles) and estimated (from the measured absorptances of PS particles) mass of nanoplastic particles, SEM images of micro- and nanofragments from teabag leachates, close-up view of NEMS-FTIR and ATR-FTIR spectra of a 500~nL teabag leachate, comparison of non-corrected and corrected ATR-FTIR spectrum of bulk nylon teabag, comparison between NEMS-FTIR and ATR-FTIR spectra regarding the spectral artifacts, experimental setup for the accelerated aging of the teabags, sampling setups, dependence of relative photothermal responsivity on the IR beam radius, FEM simulation presenting temperature distribution in a silicon nitride membrane induced by laser heating, table of characteristic peak positions (in wavenumbers) related to nanoplastic particles and nylon teabag leachates, table of values for deposited and estimated mass of polystyrene nanoparticles along with the standard deviations.

\end{footnotesize}

\enlargethispage{-1cm}

\section{REFERENCES}
\bibliographystyle{ieeetr}
\bibliography{references}

\section*{Acknowledgements}

We sincerely thank J. Smoliner, A. Lugstein and M. Sistani from the Institute of Solid State Electronics at TU Wien for letting us use their FTIR spectrometer. We gratefully acknowledge N. Barrabés Rabanal and the Research Group for Model Catalysis and Applied Catalysis at TU Wien for kindly providing access to their ATR module and supporting our measurements. We thank M. Schneider and P. Martini from the Institute of Sensor and Actuator Systems at TU Wien for acquiring the SEM images and helping with finite element method simulations, respectively.  Special thanks go to the Invisible-Light Labs team for their technical expertise and supportive encouragement, as well as to our colleagues from the MNS group at TU Wien (Institute of Sensor and Actuator Systems) for their valuable insights and collaboration. We also gratefully acknowledge G. Pfusterschmied for providing the image in Fig.~\ref{fig: Figure 1}B. We gratefully thank M. Surdu from the Extreme Environments Research Laboratory at EPFL in Switzerland for his assistance with the aerosol sampling setup.\\

\noindent\textbf{Funding:} This project received funding from the European Commission under grant agreement HORIZON-EIC-2021-TRANSITION-OPEN-101058011-NEMILIES and EIC-Transition-Booster-EMILIE-101202787, the Austria Wirtschaftsservice (AWS SEED, Deep Tech \& Physical Sciences project P2422873), the Austrian Science Fund (FWF) under project I~6086-N, and the Novo Nordisk Foundation under grant number NNF22OC0077964 - MASMONADE.\\

\noindent\textbf{Author contributions:} \\
Conceptualization: JTP, JPL, and SS. 
Methodology: JTP, JH, EŠ, NL, HB, JPL, and SS.
Investigation: JTP, JH, EŠ, NL, AG, and HB.
Supervision: AG, JPL, and SS.
Writing-original draft: JTP and EŠ.
Writing-review \& editing: JTP, JPL, and SS.\\

\noindent\textbf{Competing interests:} NL, HB, JPL, and SS are co-founders of Invisible-Light Labs GmbH who provided the analytical instrumentation used in this study. JTP, NL, HB, and JH are employed by Invisible-Light Labs GmbH. EŠ and AG declare no competing interests.\\

\vfill

\pagebreak
\setcounter{section}{0}
\setcounter{figure}{0} 
\setcounter{table}{0}
\setcounter{page}{1}
\onecolumn

\renewcommand{\baselinestretch}{1.0}

\renewcommand{\thesection}{\Alph{section}.}

\renewcommand{\thesubsection}{\arabic{subsection}}

\renewcommand{\thefootnote}{\fnsymbol{footnote}}

\fancypagestyle{firstpage}{
    \fancyhf{} 
    \renewcommand{\headrulewidth}{0pt} 
    \renewcommand{\footrulewidth}{0pt} 
}

\setlength{\headsep}{25pt} 

\titleformat{\section}[block]{\normalfont\normalsize\bfseries}{\thesection}{1em}{}
\titlespacing*{\section}{0pt}{12pt}{2pt}

\titleformat{\subsection}{\normalfont\normalsize\bfseries}{}{0pt}{}
\titlespacing*{\subsection}{0pt}{2pt}{2pt}

\titleformat{\subsubsection}{\normalfont\normalsize\bfseries\itshape}{}{0pt}{}
\titlespacing*{\subsubsection}{0pt}{4pt}{2pt}

\setlength{\columnsep}{6mm}


\title{
    \begin{center}
    \fontsize{18pt}{20pt}\selectfont Supporting Information for\\[1mm] 
        \textbf{Picogram-Level Nanoplastic Analysis with Nanoelectromechanical System Fourier Transform Infrared Spectroscopy: NEMS-FTIR}
    \end{center}
}


\author{
    \centering \fontsize{11pt}{11pt}\selectfont Jelena~Timarac--Popovi\'c\textsuperscript{1,2*}, Johannes~Hiesberger\textsuperscript{2}, Eldira~\v{S}esto\textsuperscript{1}, Niklas~Luhmann\textsuperscript{2},\\ 
    \fontsize{11pt}{11pt}\selectfont Ariane~Giesriegl\textsuperscript{1}, Hajrudin~Be\v{s}i\'c\textsuperscript{1,2}, Josiane~P.~Lafleur\textsuperscript{2}, and Silvan~Schmid\textsuperscript{1*}\footnotetext{Correspondence e--mail address: jelena.popovic@tuwien.ac.at, silvan.schmid@tuwien.ac.at} \\[2mm]
    \fontsize{10pt}{10pt}\selectfont \textsuperscript{1}\textit{TU Wien, Institute of Sensor and Actuator Systems, Gusshausstrasse 27--29, 1040 Vienna, Austria.} \\[1mm]
    \fontsize{10pt}{10pt}\selectfont \textsuperscript{2}\textit{Invisible--Light Labs GmbH, Taubstummengasse 11, 1040 Vienna, Austria.} \\[1mm]
    }

{\centering \date{\vspace{-3pt} \fontsize{10pt}{12pt}\selectfont (Dated: \today) \vspace{0pt}}\par }

\captionsetup[figure]{
    name=Figure,
    labelfont=bf,
    labelsep=period
}

\renewcommand{\thefigure}{S\arabic{figure}}

\captionsetup[table]{
    name=Table,
    labelfont=bf,
    labelsep=period
}

\renewcommand{\thetable}{S\arabic{table}}

\thispagestyle{firstpage}

\vspace{2cm}

\vfill


\clearpage

\section{Derivation of the decadic linear attenuation coefficient for sparse particle film}

In the case of sparsely distributed particles with negligible optical interactions, each particle can be considered an independent absorber. For particles much smaller than the wavelength of incident light $(R<<\lambda)$, the absorption cross section is given by \cite{baffou2013thermo}: 

\begin{equation}
    \sigma_\text{abs}(\tilde\nu) = 8 \pi^2 \tilde\nu R^3 \Im {\dfrac{\tilde n^2(\tilde\nu)-n^2_\text{m}(\tilde\nu)}{\tilde n^2(\tilde\nu)+2 n^2_\text{m}(\tilde\nu)}},
\end{equation}

\noindent where $\tilde\nu = 1/\lambda$ is the wavenumber, $\tilde n (\tilde\nu)$ the complex refractive index of the particles, and $n_m(\tilde\nu)$ the real refractive index of the surrounding medium.  

The particles sit on top of a suspended dielectric film (SiN membrane), which is optically thin compared to the wavelength $\lambda$. Hence, the refractive index of the medium is approximately that of vacuum $n_m(\tilde\nu)\approx 1$, resulting in:
\begin{equation}
    \sigma_\text{abs}(\tilde\nu) \approx 8 \pi^2 \tilde\nu R^3 \Im {\dfrac{\tilde n^2(\tilde\nu)-1}{\tilde n^2(\tilde\nu)+2 }}.
\end{equation}

The decadic volume attenuation coefficient of a particle is given by:

\begin{equation}
    \mu_{10,v}(\tilde\nu) = \dfrac{\sigma_\text{abs}(\tilde\nu)}{\ln (10)},
\end{equation}

\noindent which can be converted to the decadic linear attenuation coefficient through division by the particle volume:

\begin{equation}
    \mu_{10}(\tilde\nu)=\dfrac{\mu_{10,v}(\tilde\nu)}{\dfrac{4}{3}\pi R^3}=\dfrac{6\pi \tilde\nu}{\ln (10)}\Im {\dfrac{\tilde n^2(\tilde\nu)-1}{\tilde n^2(\tilde\nu)+2}}.
\end{equation}


\pagebreak

\begin{figure*}[h!]
\centering
\includegraphics[width=\textwidth]{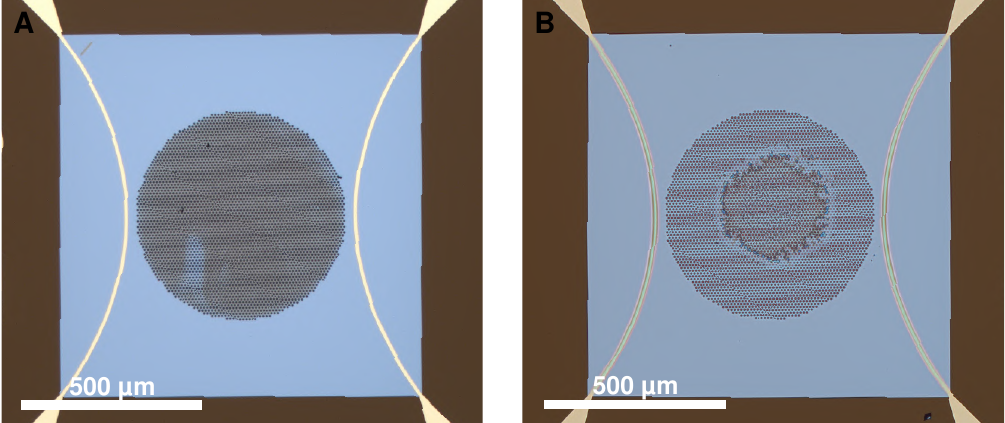}
\caption{\textbf{Sample distribution after aerosol-based deposition and drop casting in conjunction with pervaporation.} (\textbf{A}) Sample distribution obtained using the aerosol-based method of deposition of the accelerated aging teabag leachates. Aerosolized particles were collected across the entire perforated area of the membrane through inertial impaction. The image corresponds to a chip sampled for 5~min at a flow rate of 0.5~L/min with teabag leachate after 5~days of accelerated aging. (\textbf{B}) ~A dried 500~nL droplet of the teabag leachate drop casted using micropipette and dried using the pervaporation method. The pervaporation process successfully confined the analyte towards the center of the perforated membrane.}
\label{fig: Figure S_Sampled_chips}
\end{figure*}


\pagebreak

\begin{figure*}[h!]
\centering
\includegraphics[width=\textwidth]{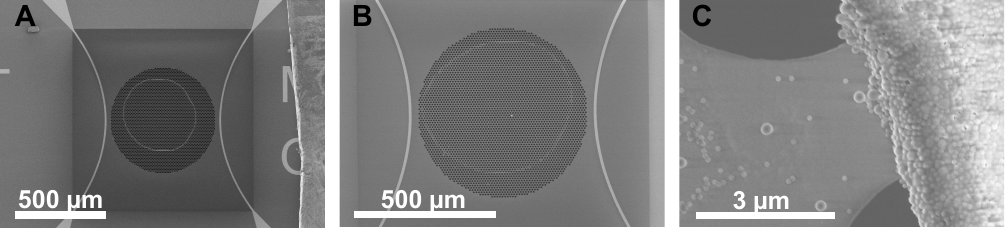}
\caption{\textbf{Sample distribution after nanodroplet dispensing.} SEM image of (\textbf{A})  1.25~ng of PP nanoparticles, (\textbf{B}) 1.25~ng of PS nanoparticles, (\textbf{C}) a PS-PP-PVC mixture mixed in a mass ratio of 1:1:1, containing 5~ng of each component, deposited onto the NEMS membrane using piezoelectric nanodroplet dispenser. The deposited samples are confined within the perforated area of the membrane following droplet drying. Images also show the coffee ring structure.}
\label{fig: Figure S_coffee ring}
\end{figure*}


\pagebreak

\begin{figure*}[h!]
\centering
\includegraphics[width=\textwidth]{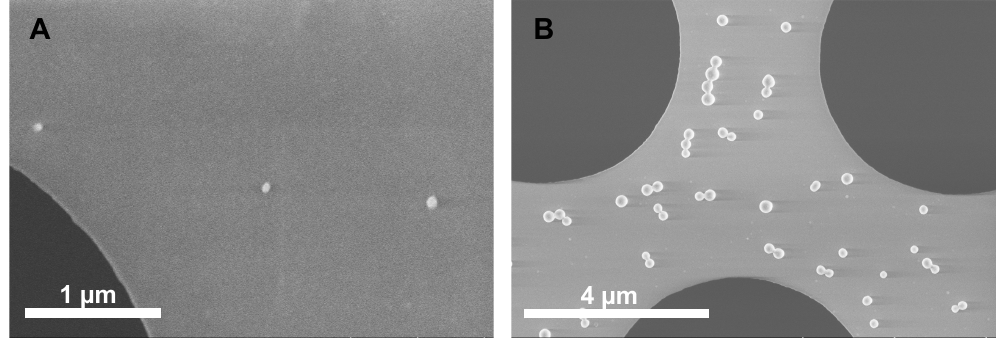}
\caption{\textbf{Nanoparticles on the perforated area of the membrane.} SEM image of (\textbf{A}) PP (nominal diameter: 54~nm), and (\textbf{B}) PVC (nominal diameter: 262~nm) nanoparticles after deposition using piezoelectric nanodroplet dispenser on the perforated area of the membrane on the NEMS chip.}
\label{fig: Figure S_SEM_PP_and_PVC}
\end{figure*}


\pagebreak

\begin{figure*}[h!]
\centering
\includegraphics[width=0.85\textwidth]{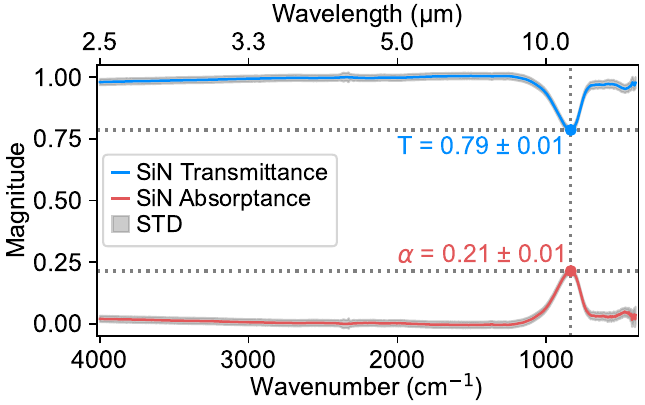}
\caption{\textbf{IR spectral features of SiN membranes.} Average transmittance and absorptance spectra of 12 individual empty NEMS chips composed of 50~nm thick low-stress SiN. The shaded regions represent the standard deviations associated with each spectrum. The dashed vertical line at 835~$\text{cm}^{-1}$ marks the characteristic SiN peak, where the transmittance and absorptance values, along with their standard deviations, were evaluated.}
\label{fig: Figure S_T and alpha of SiN}
\end{figure*}


\pagebreak

\begin{figure*}[h!]
\centering
\includegraphics[width=0.85\textwidth]{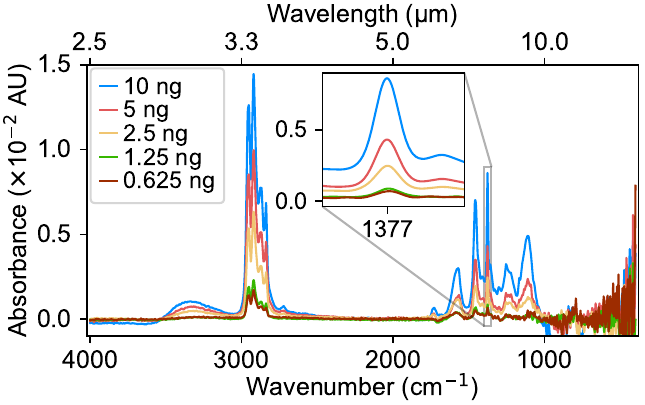}
\caption{\textbf{Characterization of PP nanoplastics.} NEMS-FTIR spectra of PP nanoplastics deposited on NEMS chips at varying mass loads. The inset highlights the 1377~cm$^{-1}$ peak, which was used to construct the calibration curve and determine the LoD.}
\label{fig: Figure S_PP NPs}
\end{figure*}


\pagebreak

\begin{figure*}[h!]
\centering
\includegraphics[width=0.85\textwidth]{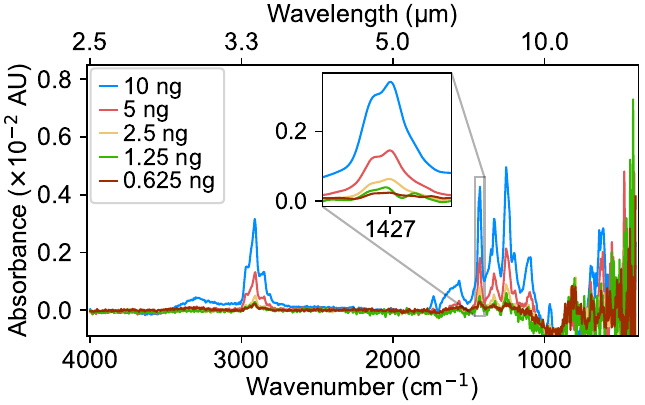}
\caption{\textbf{Characterization of PVC nanoplastics.} NEMS-FTIR spectra of PVC nanoplastics deposited on NEMS chips at varying mass loads. The inset highlights the 1427~cm$^{-1}$ peak, which was used to construct the calibration curve and determine the LoD.}
\label{fig: Figure S_PVC NPs}
\end{figure*}


\pagebreak

\begin{figure*}[h!]
\centering
\includegraphics[width=0.85\textwidth]{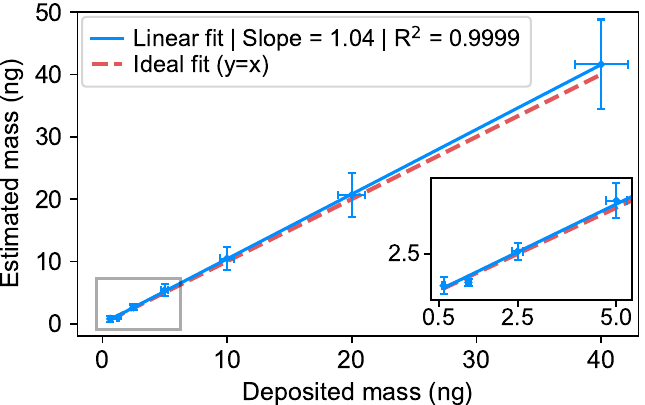}
\caption{\textbf{Quantitative analysis.} Correlation between the mass of PS nanoplastic particles deposited on the NEMS chip and the estimated mass derived from the measured NEMS-FTIR spectra after conversion to optical absorbance (N=3). The inset highlights the lower mass range for improved visibility. Error bars represent propagated uncertainties from both the measurement and estimation processes.}
\label{fig: Figure S_Deposited_vs_estimated_mass}
\end{figure*}


\pagebreak

\begin{figure*}[h!]
\centering
\includegraphics[width=\textwidth]{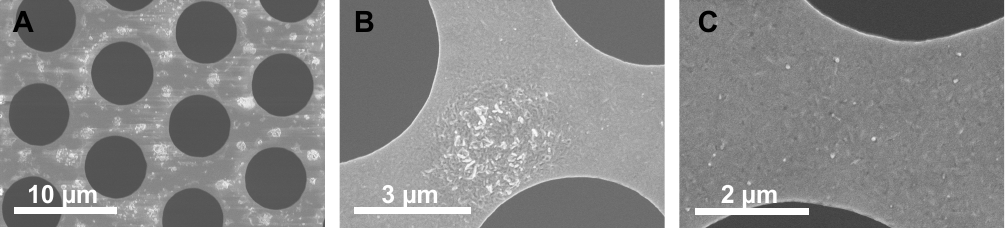}
\caption{\textbf{Micro- and nanofragments released by teabags during accelerated aging.}~(\textbf{A-C})~Magnified views of the NEMS membrane resonator sampled with water in which a teabag had been soaked for 15 days and exposed to environmental stress.}
\label{fig: Figure S_SEM_teabag_acc aging}
\end{figure*}


\pagebreak

\begin{figure*}[h!]
\centering
\includegraphics[width=0.85\textwidth]{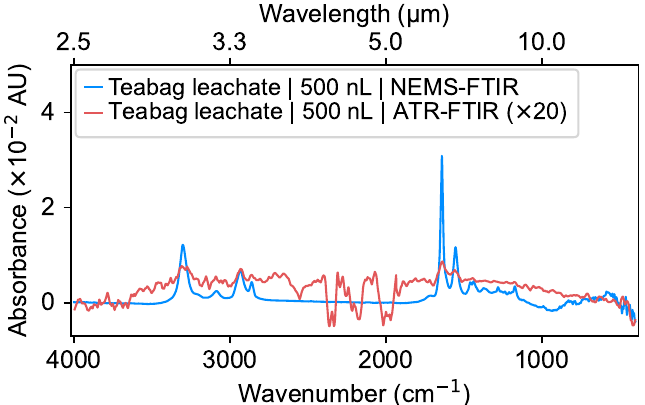}
\caption{\textbf{Comparison of NEMS-FTIR and ATR-FTIR spectra of a 500~nL teabag leachate.} The ATR-FTIR spectrum of 500~nL of teabag leachate is multiplied by a factor of 20 with respect to the NEMS-FTIR spectrum of 500~nL of the same leachate for improved visibility, revealing only weak indications of the most prominent IR peaks related to nylon.}
\label{fig: Figure S_500nL TBL_close-up}
\end{figure*}


\pagebreak

\begin{figure*}[h!]
\centering
\includegraphics[width=\textwidth]{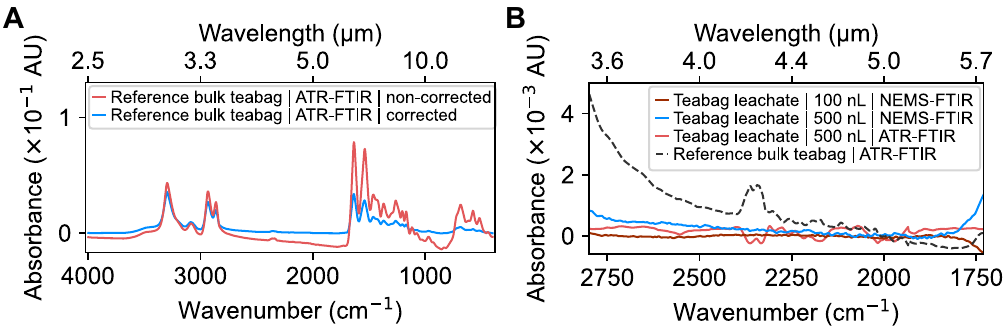}
\caption{\textbf{Correction effects and spectral artifacts in ATR-FTIR.}~(\textbf{A})~Comparison of non-corrected and corrected ATR-FTIR spectrum of bulk nylon teabag (reference). Corrections applied include "Extended ATR correction" (standard mode) and "Baseline correction" using OPUS software. The figure demonstrates how the applied corrections alter the relative intensities of spectral peaks. (\textbf{B})~Comparison of NEMS-FTIR spectra of 100~nL and 500~nL nylon teabag leachates with ATR-FTIR spectra of 500~nL of the same leachate and the bulk nylon teabag in the 2800–1730~cm$^{-1}$ region. CO$_2$ peaks and diamond phonon bands are visible in the ATR-FTIR spectra, whereas these spectral artifacts are absent in the NEMS-FTIR spectra.}
\label{fig: Figure S_ATR_artifacts_and_correction_effect}
\end{figure*}


\pagebreak

\begin{figure*}[h!]
\centering
\includegraphics[width=0.5\textwidth]{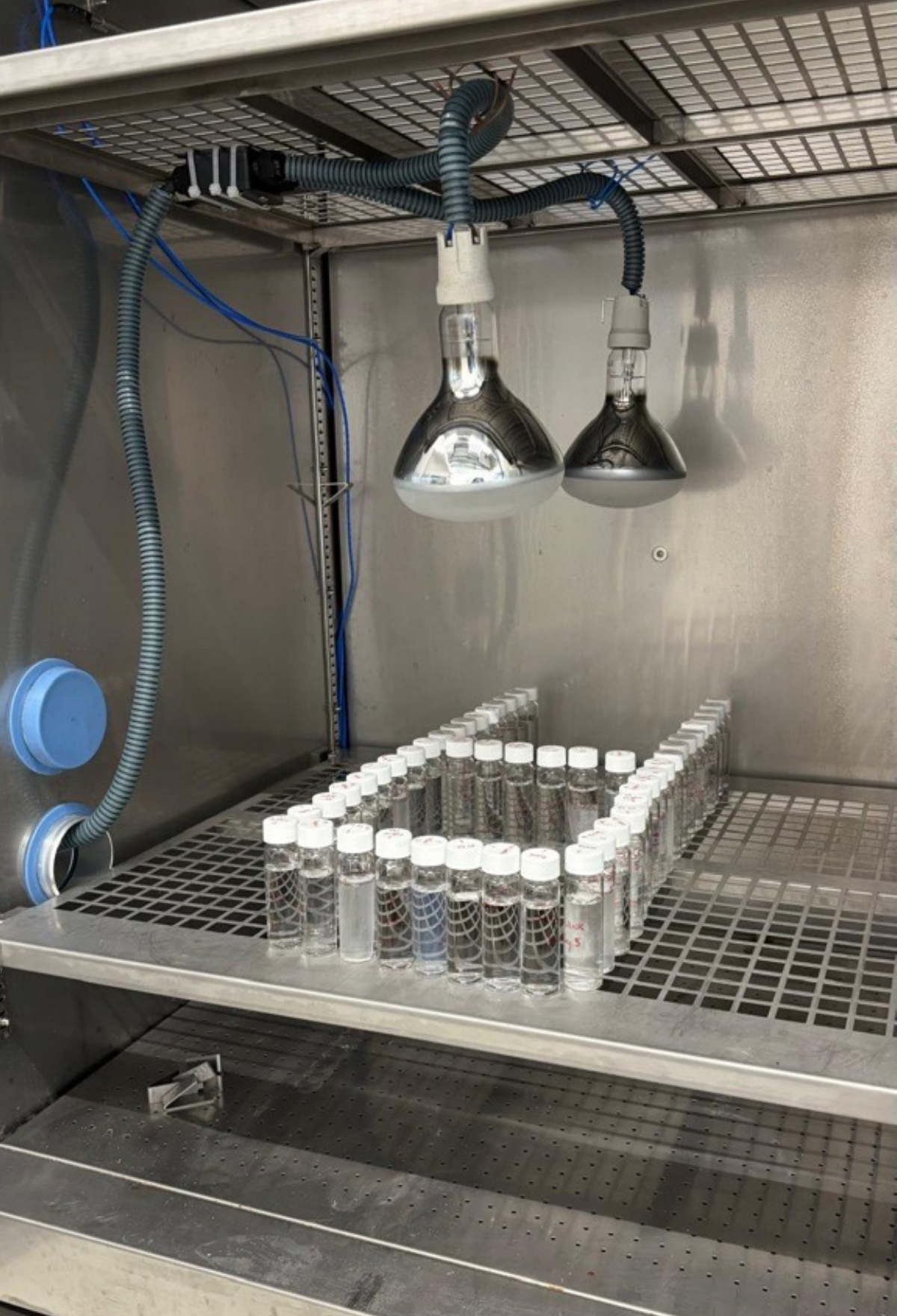}
\caption{\textbf{Experimental setup for the accelerated aging of teabags.} The setup includes a controlled environment chamber for simulating environmental stress through cyclic exposure to elevated temperature and UV radiation, with teabags immersed in water for the duration of the experiment.}
\label{fig: Figure S_climate chamber}
\end{figure*}


\pagebreak

\begin{figure*}[h!]
\centering
\includegraphics[width=\textwidth]{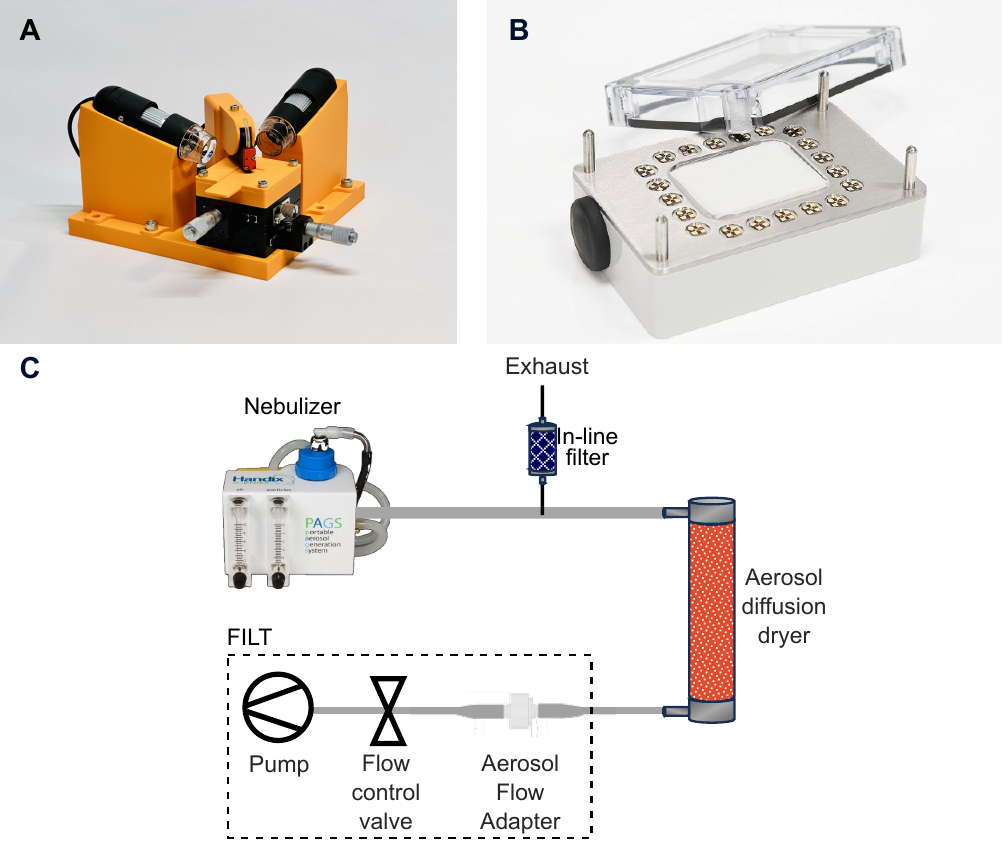}
\caption{\textbf{Sampling setups.} (\textbf{A}) The nanodroplet dispensing set-up consisted of the piezoelectric nanodroplet dispenser (PIPEJET nanoDispenser, BioFluidix-Hamilton Freiburg, Germany) equipped with two cameras (USB Mikroskop, Prod. No. TO-5139591, TOOLCRAFT, Germany) and the XY stage (XY Axis Manual Displacement Platform, Prod. No. B07TQRWQH4, Amazon, Germany) mounted using a custom 3D-printed scaffold. The set-up enabled precise positioning of the NEMS chips beneath the dispensing capillary for accurate droplet deposition onto the central region of the perforated membrane area. (\textbf{B}) The Drop Casting Accessory (Invisible-Light Labs GmbH, Austria) for liquid samples. The accessory allows for precise sample deposition and controlled drying under a humidity gradient \textit{via} pervaporation, ensuring consistent localization of the sample within the perforated membrane area. (\textbf{C}) Schematic representation of the setup, comprising a nebulizer (Portable Aerosol Generation System, PAGS, Handix Scientific Inc., CO, USA) for aerosolizing the liquid into fine droplets, a diffusion dryer (DDU, 570/L, Topas GmbH, Germany), and a NEMS chip holder (Aerosol Flow Adapter, Invisible-Light Labs GmbH, Austria) containing the NEMS chip. The chip holder was placed in an eight-channel filter sampler (FILT, Brechtel Manufacturing Inc., CA, USA), which includes a built-in pump. Sampling is based on the inertial impaction deposition mechanism, enabling sample deposition across the entire perforated membrane area.}
\label{fig: Figure S_Sampling setups}
\end{figure*}


\pagebreak

\begin{figure*}[h!]
\centering
\includegraphics[width=0.85\textwidth]{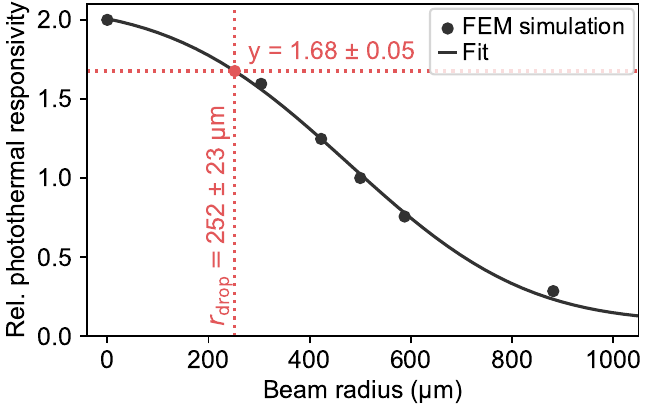}
\caption{\textbf{Dependence of relative photothermal responsivity on the IR beam radius.} Data obtained through FEM simulations of a 1x1~mm$^{2}$ low-stress SiN membrane. The graph highlights how varying the beam radius affects the responsivity of the NEMS membrane due to spatial heat distribution.}
\label{fig: Figure S_responsivity vs. beam radius}
\end{figure*}


\pagebreak

\begin{figure*}[h!]
\centering
\includegraphics[width=\textwidth]{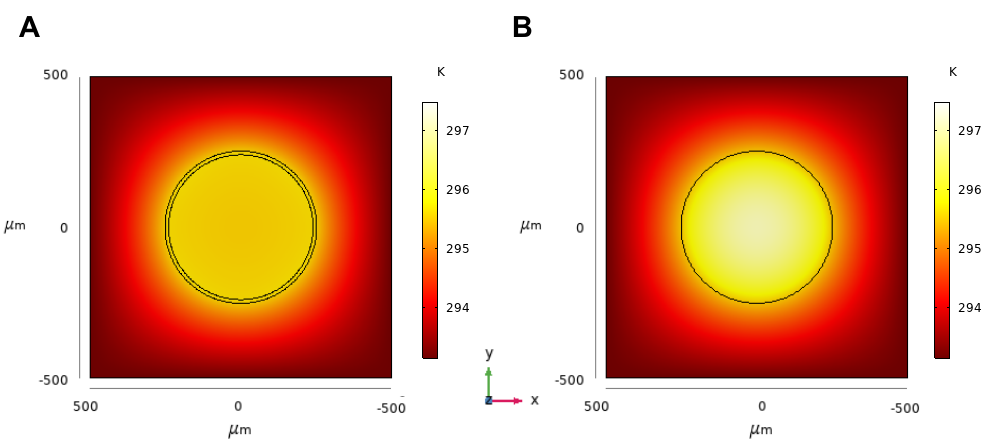}
\caption{\textbf{Temperature distribution in a SiN membrane induced by laser heating.} \textbf{(A)} Simulated temperature distribution on a 1×1~mm$^2$ SiN membrane resulting from a ring-shaped heat source (diameter 252~$\mu$m), mimicking a donut-shaped laser beam. \textbf{(B)} Temperature distribution on the same membrane with a disk-shaped heat source of the same diameter. The difference in photothermal optomechanical response between these distributions is approximately 3\%.}
\label{fig: Figure S_FEM simulations}
\end{figure*}


\clearpage
\setlength{\LTcapwidth}{\linewidth}

\begin{longtable}{p{2cm} p{2cm} p{9.5cm} p{2cm}}
\caption{\color{blue}\textbf{Characteristic IR absorption bands of plastics.}
Overview of the main IR absorption peaks identified in the NEMS-FTIR spectra for PS, PP, PVC, and PA particles, together with corresponding band assignments. The cited references originate from transmission and ATR-FTIR measurements, demonstrating direct comparison between NEMS-FTIR and established IR spectroscopy data.}
\label{tab:NPpeaks}
\\[0.8em]
\hline
\\[-0.9em]
\textbf{\parbox{2cm}{\raggedright Type of\\plastics}} & 
\textbf{\parbox{2cm}{\raggedright Peak position\\(cm$^{-1}$)}} & 
\textbf{\parbox{9.5cm}{\raggedright Assignment}} & 
\textbf{\parbox{2cm}{\raggedright References}} \\[1.2em]
\hline
\\[-0.9em]
\endfirsthead

\hline
\\[-0.9em]
\textbf{\parbox{2cm}{\raggedright Type of\\plastics}} & 
\textbf{\parbox{2cm}{\raggedright Peak position\\(cm$^{-1}$)}} & 
\textbf{\parbox{6cm}{\raggedright Assignment}} & 
\textbf{\parbox{2cm}{\raggedright References}} \\[1.2em]
\hline
\\[-0.9em]
\endhead

\multirow{16}{*}{PS} 
& 700  & Aromatic ring bending & \cite{Smith2021PSandPPpeaks, Wu2019PSpeaks, NIST2025PSpeaks, Zajac2022PSpeaks} \\
& 753  & Aromatic out-of-plane C–H bending & \cite{Smith2021PSandPPpeaks, Wu2019PSpeaks, NIST2025PSpeaks, Zajac2022PSpeaks} \\
& 1028 & C–H in-plane bending of the phenyl ring & \cite{Wu2019PSpeaks, NIST2025PSpeaks, Boke2022O-PTIR_PSPPPVCpeaks} \\
& 1452 & C–H bending (scissoring, aliphatic backbone) & \cite{Wu2019PSpeaks, NIST2025PSpeaks, Boke2022O-PTIR_PSPPPVCpeaks, Zajac2022PSpeaks} \\
& 1493 & \multirow{3}{7cm}{Aromatic C–C stretching (ring vibration)} & \cite{Smith2021PSandPPpeaks, Wu2019PSpeaks, NIST2025PSpeaks, Boke2022O-PTIR_PSPPPVCpeaks, Zajac2022PSpeaks} \\
& 1583 &  & \cite{Wu2019PSpeaks, NIST2025PSpeaks} \\
& 1601 &  & \cite{Smith2021PSandPPpeaks, Wu2019PSpeaks, NIST2025PSpeaks, Boke2022O-PTIR_PSPPPVCpeaks, Zajac2022PSpeaks} \\
& 1745 & \multirow{4}{7cm}{Aromatic overtone/combination band (benzene fingers)} & \cite{Smith2021PSandPPpeaks, NIST2025PSpeaks} \\
& 1800 &  & \cite{Smith2021PSandPPpeaks, Wu2019PSpeaks, NIST2025PSpeaks} \\
& 1868 &  & \cite{Smith2021PSandPPpeaks, Wu2019PSpeaks, NIST2025PSpeaks} \\
& 1943 &  & \cite{Smith2021PSandPPpeaks, Wu2019PSpeaks, NIST2025PSpeaks} \\
& 2849 & \multirow{2}{7cm}{Aliphatic C–H stretching (CH$_2$ asymmetric and symmetric)} & \cite{Smith2021PSandPPpeaks, Wu2019PSpeaks, NIST2025PSpeaks, Zajac2022PSpeaks} \\
& 2923 &  & \cite{Smith2021PSandPPpeaks, Wu2019PSpeaks, NIST2025PSpeaks, Zajac2022PSpeaks} \\
& 3026 & \multirow{3}{7cm}{Aromatic C–H stretching} & \cite{Smith2021PSandPPpeaks, Wu2019PSpeaks, NIST2025PSpeaks, Zajac2022PSpeaks} \\
& 3060 &  & \cite{Smith2021PSandPPpeaks, Wu2019PSpeaks, NIST2025PSpeaks} \\
& 3081 &  & \cite{Smith2021PSandPPpeaks, Wu2019PSpeaks, NIST2025PSpeaks} \\
\hline
\\[-0.9em]

\multirow{11}{*}{PP} 
& 1115 & C–C chain stretching + CH$_3$ rocking + CH$_2$ wagging + CH twisting + CH bending & \cite{karacan2011PPpeaks} \\
& 1167 & C–H bending + C–C chain stretching + CH$_3$ rocking & \cite{Tagg2017PPPVCPApeaks, Boke2022O-PTIR_PSPPPVCpeaks, karacan2011PPpeaks} \\
& 1220 & CH$_2$ twisting + CH bending + C–C chain stretching & \cite{karacan2011PPpeaks} \\
& 1254 & CH bending + CH$_2$ twisting + CH$_3$ rocking & \cite{karacan2011PPpeaks} \\
& 1377 & CH$_3$ bending, symmetric (umbrella mode) & \cite{Tagg2017PPPVCPApeaks, Boke2022O-PTIR_PSPPPVCpeaks, Fang2012PPpeaks} \\
& 1459 & CH$_3$ bending, symmetric & \cite{Tagg2017PPPVCPApeaks, Boke2022O-PTIR_PSPPPVCpeaks, Fang2012PPpeaks} \\
& 2839 & CH$_2$ stretching, symmetric & \cite{Smith2021PSandPPpeaks, Tagg2017PPPVCPApeaks} \\
& 2868 & CH$_3$ stretching, symmetric & \cite{Smith2021PSandPPpeaks, Tagg2017PPPVCPApeaks, Fang2012PPpeaks} \\
& 2920 & CH$_2$ stretching, asymmetric & \cite{Smith2021PSandPPpeaks, Tagg2017PPPVCPApeaks, Fang2012PPpeaks} \\
& 2952 & CH$_3$ stretching, asymmetric & \cite{Smith2021PSandPPpeaks, Tagg2017PPPVCPApeaks, Fang2012PPpeaks} \\
\hline
\\[-0.9em]

\multirow{13}{*}{PVC} 
& 614 & \multirow{3}{*}{C-Cl stretching} & \cite{beltran1997PVCpeaks, Boke2022O-PTIR_PSPPPVCpeaks} \\
& 637 &  & \cite{beltran1997PVCpeaks, Boke2022O-PTIR_PSPPPVCpeaks} \\
& 692 &  & \cite{beltran1997PVCpeaks, stromberg1958PVCpeaks, Boke2022O-PTIR_PSPPPVCpeaks} \\
& 970 & CH$_2$ rocking & \cite{beltran1997PVCpeaks, stromberg1958PVCpeaks, Tagg2017PPPVCPApeaks, Boke2022O-PTIR_PSPPPVCpeaks} \\
& 1100 & C-C stretching & \cite{beltran1997PVCpeaks, stromberg1958PVCpeaks, Tagg2017PPPVCPApeaks} \\
& 1254 & 
{C-H bending, in-phase} 
& \cite{beltran1997PVCpeaks, stromberg1958PVCpeaks, Tagg2017PPPVCPApeaks} \\
& 1334 & C-H bending, out-of-phase & \cite{beltran1997PVCpeaks, Fawzi2025PVCpeaks, stromberg1958PVCpeaks, Tagg2017PPPVCPApeaks} \\
& 1427 & CH$_2$ bending, in-phase & \cite{beltran1997PVCpeaks, stromberg1958PVCpeaks, Tagg2017PPPVCPApeaks} \\
& 1625 & C=C stretching & \cite{Fawzi2025PVCpeaks} \\
& 2849 & CH$_2$ stretching, symmetrical, in-phase & \cite{stromberg1958PVCpeaks} \\
& 2910 & C-H stretching of CH$_2$, asymmetrical, in-phase & \cite{beltran1997PVCpeaks, stromberg1958PVCpeaks, Tagg2017PPPVCPApeaks} \\
& 2971 & C-H stretching of CHCl, out-of-phase & \cite{beltran1997PVCpeaks, stromberg1958PVCpeaks} \\
\\[-0.9em]

\multirow{17}{*}{PA} 
& 575 & C=O out-of-plane vibration (amide VI) & \cite{kang2022PApeaks, Tummino2023PApeaks} \\
& 990 & C-CO stretching (amide IV) & \cite{kang2022PApeaks, Tummino2023PApeaks, gonccalves2007PApeaksdegradation} \\
& 1116 & C-N stretching & \cite{Shurvell2001PApeaks, Smith2023PApeaks} \\
& 1170 & C-C stretching & \cite{kang2022PApeaks} \\
& 1210 & CH$_2$ wagging or twisting & \cite{kang2022PApeaks} \\
& 1235 & C-N stretching & \cite{Smith2023PApeaks} \\
& 1282 & C-N stretching + CH$_2$ twisting & \cite{Smith2023PApeaks, kang2022PApeaks} \\
& 1371 & CH$_2$ wagging + C-N-H deformation & \cite{kang2022PApeaks, gonccalves2007PApeaksdegradation, Tummino2023PApeaks} \\
& 1437 & CH$_2$ (Gauche bonds) deformation & \cite{kang2022PApeaks} \\
& 1460 & CH$_2$ (adjacent to N-H group) deformation & \cite{kang2022PApeaks, Tummino2023PApeaks} \\
& 1553 & N-H in-plane bending + C-N stretching (amide II) & \cite{Smith2023PApeaks, kang2022PApeaks, Tummino2023PApeaks} \\
& 1642 & C=O stretching (amide I) & \cite{Smith2023PApeaks, kang2022PApeaks, Tummino2023PApeaks} \\
& 1720 & Imide group; carboxylic acid C=O stretching mode & \cite{gonccalves2007PApeaksdegradation, Zimudzi2018PApeakscarboxylicacid, OkambaDiogo2014PApeaksimidecarboxylicacid} \\
& 2860 & C-H stretching, symmetrical & \cite{kang2022PApeaks, Tummino2023PApeaks} \\
& 2930 & C-H stretching, asymmetrical & \cite{kang2022PApeaks, Tummino2023PApeaks} \\
& 3087 & C-N stretching + overtone of N-H in-plane bending & \cite{Smith2023PApeaks, kang2022PApeaks, gonccalves2007PApeaksdegradation, Tummino2023PApeaks} \\
& 3300 & N-H stretching & \cite{Smith2023PApeaks, kang2022PApeaks, Tagg2017PPPVCPApeaks, Tummino2023PApeaks, Shurvell2001PApeaks} \\
\hline
\end{longtable}
\clearpage


\pagebreak

\begin{table}[h!]
    \centering
    \captionsetup{width=\textwidth}
    \caption{\textbf{Deposited and estimated mass of PS NPs.} Quantitative comparison of the deposited and estimated mass values for PS nanoparticles on NEMS chips, accompanied by their respective standard deviations. Deposited masses were derived from the dispersion concentrations and droplet volume, with uncertainties propagated from sample preparation (micropipette) and deposition tools (piezoelectric nanodroplet dispenser). Estimated masses were calculated using Eq.~(6). The main source of uncertainty in estimated mass originates from variability in the measured absorptance at 1452~cm$^{-1}$.}
    \label{tab:Deposited vs. estimated mass}
    \resizebox{\linewidth}{!}{%
    \begin{tabular}{cccccc}
    \hline
    \\[-0.7em]
        \textbf{\parbox{3.7cm}{\centering Deposited mass\\(ng)}} & \textbf{\parbox{3.7cm}{\centering SD\\(ng)}} & \textbf{\parbox{3.7cm}{\centering Estimated mass\\(ng)}} & \textbf{\parbox{3.7cm}{\centering SD\\(ng)}} \\[0.7em]
        \hline
        \\[-0.9em]
        0.62 & 0.03 & 0.8 &  0.5 \\
        1.2 & 0.1 & 1.0 &  0.2 \\
        2.5 & 0.1 & 2.6 &  0.5 \\ 
        5.0 & 0.3 & 5 & 1 \\ 
        10.0 & 0.5 & 10 & 2 \\ 
        20 & 1 & 21 & 4 \\ 
        40 & 2 & 42 & 7 \\ [0.1em]
        \hline
    \end{tabular}%
    }
\end{table}


\end{document}